\documentstyle[12pt,psfig]{article}
\def\labell#1{\label{#1}}

\def\Tr{\mathop{\rm Tr}}
\def\s{s}
\def\t{t}
\def\u{s'}
\def\v{t'}

\def\X{Q}
\def\Y{R}

\begin{document}
\renewcommand{\thefootnote}{\fnsymbol{footnote}}

\begin{center}
{\Large  {\bf A geometric approach to free variable loop equations
in discretized theories of 2D gravity}}
\footnote[1]{This work was supported in part by 
the National Science Foundation under grants PHY/9200687 and PHY/9108311,
by the U.S. Department of Energy (D.O.E.) under cooperative agreement 
DE-FC02-94ER40818, by the divisions of 
applied mathematics of the D.O.E. under contracts DE-FG02-88ER25065
and DE-FG02-88ER25065, and by the European Community Human Capital Mobility 
programme.}\\  
\vspace{0.2in}
Sean M. Carroll,$^{(1)}$ Miguel E. Ortiz$^{(2)}$\footnote[2]{Address after
January 1996: Blackett Laboratory, Imperial College, Prince Consort Road,
London SW7 2BZ, UK.} 
and 
Washington Taylor IV$^{(1)}$\footnote[3]{Address after January 1996:
Department of Physics, Joseph Henry Laboratories, Princeton
University, Princeton NJ 08544.}\\
\vspace{0.1in}
{\it $^{(1)}$Center for Theoretical Physics, Laboratory for
Nuclear Science}\\
{\it and Department of Physics}\\
{\it Massachusetts Institute of Technology}\\
{\it Cambridge, Massachusetts\quad 02139}\\
{\it email: carroll@ctp.mit.edu, wati@mit.edu}\\
\vspace{0.1in}
{\it $^{(2)}$Institute of Cosmology, Department of Physics and Astronomy,}\\
{\it Tufts University, Medford, MA 02155}\\
{\it and}\\
{\it Instituut voor Theoretische Fysika, Universiteit Utrecht}\\
{\it Princetonplein 5, 3508 TA Utrecht, The Netherlands}\\
{\it email: ortiz@fys.ruu.nl}\\
\end{center}
\begin{abstract}
We present a self-contained analysis of theories of discrete 2D
gravity coupled to matter, using geometric methods to derive equations for
generating functions in terms of free (noncommuting) variables.  For the
class of discrete gravity theories which correspond to matrix models, our
method is a generalization of the technique of Schwinger-Dyson 
equations and is closely related to recent work describing the master field
in terms of noncommuting variables; the important differences are that we
derive a single equation for the generating function using purely graphical
arguments, and that the approach is applicable to a broader class of
theories than those described by matrix models.  Several example
applications are given here, including theories of gravity coupled to a
single Ising spin ($c = 1/2$), multiple Ising spins ($c = k/2$), a general
class of two-matrix models which includes the Ising theory and its dual,
the three-state Potts model, and a dually weighted graph model which does
not admit a simple description in terms of matrix models.

\end{abstract}

\renewcommand{\thefootnote}{\arabic{footnote}}
\vfill
\centerline{CTP~\# 2465 \hfill October 1995}
\centerline{THU-95/28\hfill}
\centerline{hep-th/9510199 \hfill}
\eject
\section{Introduction}

A major goal of theoretical physics is the construction of a
self-consistent quantum theory of gravity coupled to matter in  four
dimensions.  Because such a theory has proven to be very difficult to
construct, it is interesting to consider toy models of gravity coupled to
matter in lower dimensions, in the hope that such simplified theories can 
provide information about the nature of quantum gravity.  In
particular, it has been possible to construct consistent quantum theories
of (Euclidean) gravity in two dimensions coupled to conformal matter.
These theories have been the subject of considerable interest over the last
decade, partly because they are the simplest available theories of gravity
coupled to matter and partly because of the relevance of these models to
string theory.  Two dimensional gravity theories of this type can be
defined either through the continuum Liouville theory approach
\cite{kpz,david3,diskawai}, or through
discrete models of dynamically triangulated gravity (for reviews
see \cite{ajw,david0,ginmoore}).

The study of discretized two-dimensional gravity theories has a long
history \cite{regge,weingarten}.  A major breakthrough occurred
when it was realized that a summation over ``triangulations'' of a
2-manifold by equilateral polygons (not necessarily triangles) can be
described by a zero-dimensional field theory, or matrix model
[9-14].
By writing a discretized 2D gravity theory as an
integral over hermitian matrices, it becomes possible to use analytic
methods to calculate correlation functions in these models, and to show
that in an appropriate continuum limit these models appear to correspond to
continuum theories of Liouville gravity coupled to conformal matter field
(for a review see \cite{dgz}).  
The matrix model technology which was developed also made it
possible to sum the partition function for these models over Riemann
surfaces of all genera, giving nonperturbative results for many theories of
this type \cite{gm,ds,brkaz}.

One approach which has been used to analyze matrix model theories
involves the use of Schwinger-Dyson equations \cite{migdal,kaz2,david2}.
These equations, derived by demanding the invariance of the matrix
model correlation functions under infinitesimal changes of variables,
may be interpreted geometrically as ``loop equations'' describing the
deformation of two-dimensional geometries.  The Schwinger-Dyson/loop
equations are a powerful tool for analyzing the algebraic structure of
correlation functions, although it has proven difficult to find
solutions in more complex theories.  Recently, the Schwinger-Dyson
equations for matrix models have been used to discuss  $c < 1$ string
field theories
[22-31].

In this paper we generalize the Schwinger-Dyson technique, developing
a systematic method for deriving a single equation satisfied by the
generating function in a very general class of discrete 2D gravity
theories.  Our approach is based on the combinatorial methods pioneered
by Tutte \cite{tutte}, who found an algebraic description of one class of
triangulations by writing a discrete recursion relation for the number
of triangulations of a 2-manifold with a fixed number of boundary
edges (see also \cite{bc,knn}).  We show how this type of recursion
relation can be derived in the presence of discrete data defined on
the triangulation, representing matter fields in the continuum.  Using
this technique we are able to describe a variety of theories, including
theories such as the recently described dually
weighted graph models \cite{dfi,ksw} which do not always admit a simple
description in terms of matrix models.  For matrix model theories,
this formalism can be used to describe generating functions for
correlation functions with a general class of boundary conditions,
which can lead to results which cannot be easily
obtained through the matrix model formalism.

In theories with matter, the generating function for the disk
amplitude is a formal sum over all possible matter configurations on
the boundary of the triangulations.  We encode boundary data for the 
matter fields as a string of free (noncommuting) variables
associated with the matter configuration on the edges or vertices of
the boundary.  In this way the generating function is an element of
the free algebra generated by the variables representing different
matter states.  One of the principal results presented in this paper
is the derivation of a ``generating equation,'' satisfied by the
generating function.  This generating equation is closely related to
the master equation which has recently been discussed by a variety of
authors 
[37-42].
The generating equation describes the
effect of removing a marked edge from a triangulation, which results
in one of two basic geometric ``moves'' -- the removal of the triangle
attached to the edge or, if the edge is connected directly to another
boundary edge, the removal of that pair of boundary edges.  In fact,
the correspondence between these moves and the generating equation is
sufficiently precise that the equation may be written down directly
upon consideration of the possible effects of the moves, without going
through the intermediate step of an explicit recursion equation.

Once we have a single generating equation in terms of noncommuting
variables for the correlation functions in the theory, the next step is to
try to find a subset of correlation functions for which a closed system of
equations can be written in terms of commuting variables.  This is
analogous to the technique
used in Ref. \cite{staudacher} to find the cubic equation satisfied by the
generating function of homogeneous correlation functions in the $c = 1/2$
matrix model which describes an Ising matter field coupled to 2D gravity;
a description of this calculation in terms of noncommuting variables was
given in Ref. \cite{dl}.  From a polynomial equation of this type it is
straightforward if somewhat tedious to derive information about the scaling
behaviour of correlation functions in the continuum limit.

A principal focus of this paper is the development of a systematic
approach to finding and solving closed sets of equations in commuting
variables starting from the noncommuting generating equation.  We
give several examples of polynomial equations which can be derived in
this fashion.  In particular, we explicitly solve a 1-parameter family
of two-matrix models which may be thought of as the Ising model
coupled to gravity with a boundary magnetic field.  This family of
models includes both the Ising theory (in which Ising spins are located
on the faces of the triangles) and its dual model (in which the spins are
located on the vertices of the triangulation).  The solution reproduces the 
cubic equations for the Ising and dual theories, and shows that the  
intermediate models satisfy a quartic.  We also describe the
generating equations for the 3-state Potts model and the $2^k$-matrix
model with $c = k/2$ which describes $k$ Ising spins coupled to 2D
gravity; in both of these examples we are unable to find a closed
system of equations in commuting variables, and we discuss the
obstacles encountered in this approach.

We do not discuss the continuum limit of any of the polynomial equations
that are derived in this paper. All the equations are at most quartic in a
single commuting generating function, and contain a finite number of
functions that correspond to boundary data for the generating function.
The continuum limit is given by the expansion of the parameters of the
theory around their critical values, as governed by the appropriate
polynomial equation; for an example see \cite{kkmw}. The continuum limit
of some of the more interesting models considered in this paper will be  
discussed in Refs. \cite{cot2,cot3}.

The approach we have taken in this paper is to develop the general
formalism in a piecewise fashion, introducing each additional
complication in the context of a particular model where extra
structure is present.  The outline of the paper is as follows: in
Section 2 we describe the geometric approach used throughout the paper
by examining in detail the generating equation for pure triangulated
2D gravity, which can be written in terms of a single variable that
encodes boundary information.  In Section 3, we consider gravity
coupled to a single Ising spin, and associate boundary configurations
with words in a free algebra generated by noncommuting variables.  We
derive the cubic equation governing homogeneous boundary   
configurations in this model.  We also give a generating equation for
a more general class of boundary variables in this model which
incorporates information about nearest-neighbor spin pairs.  In 
Section 4 we study the Ising theory on the dual lattice, and derive the
cubic equation for the homogeneous disk amplitude in this formulation
of the model.  In
Section 5 we find the generating equation for a general two-matrix
model with cubic interactions, which includes the Ising theory, dual
Ising theory, and the 1-parameter family of models describing the
Ising theory with boundary magnetic field.  We also outline the
derivation of a quartic equation for the homogeneous disk amplitude in
these models.  Some of the results mentioned in this section will be
described in greater detail in a separate paper \cite{cot3}.  In
Section 6 we examine the 3-state Potts model, which serves as an
example of the obstacles faced by this kind of approach.  In Section
7, we describe the generating equation for the theory of triangulated
gravity coupled to multiple Ising spins ($c = k/2$).  In Section 8, we
consider a more general class of discrete gravity models, and derive
the generating equation for a simple dually weighted graph model.
Section 9 contains concluding remarks.

\section{Pure Gravity}
\label{sec:puregravity}

In this section we consider pure 2D gravity in the absence of matter
fields; this corresponds to the generic one-matrix model.  We use this
simple model to present an introduction to the geometrical techniques
that are applied to solve more complex models in subsequent sections.
We derive the well known equation for the disk amplitude generating
function, in the case of cubic interactions \cite{BIPZ}.  Although the
generating function in this model is written in terms of a single
commuting variable, we show how this result may be written in a form
that generalizes to free variables.  Finally, we derive a more
elaborate system of generating equations for higher-genus amplitudes.

\subsection{Review of model}

The partition function for pure 2D gravity is formally defined as
\begin{equation}
Z = \sum_{h} \left( \int {{\cal D}_{\cal M}\left[g_{\mu\nu}\right]\over
{\rm Diff}_{\cal M}} e^{-S_E[g_{\mu\nu}]}\right) \ , 
\labell{eq:cont}
\end{equation}
where the genus $h$  of the orientable 2-manifold ${\cal M}$  is
summed over and
\begin{equation}
  S_E[g_{\mu\nu}]={1\over 16\pi G}\int d^2x\sqrt{g}\left(-R+2\Lambda\right)\ .
  \labell{eq:eh}
\end{equation}
A discretized version of
this theory can be constructed by summing over all triangulated
2-manifolds.  This discrete theory has a partition function
\begin{equation}
  Z = \sum_{\Delta} \frac{1}{S (\Delta)}  N^{\chi (\Delta)} 
  g^{n (\Delta)}\ ,
  \labell{eq:pgpf}
\end{equation}
where $g$ and $N$ are coupling constants that replace $\Lambda$ and $G$,
$\chi (\Delta)$ is the Euler character of the triangulation $\Delta$, $S
(\Delta)$ is the symmetry factor of $\Delta$, and $n(\Delta)$ is the number
of triangles in $\Delta$. The matrix model
expression for this partition function is given by
\begin{equation}
  Z = \int {\rm D} U \;\exp \left(-N\left[\frac{1}{2} {\rm Tr}\; U^2
  - \frac{g}{3} {\rm Tr}\; U^3 \right]\right) \ ,
  \labell{eq:pgmmpf}
\end{equation}
where $U$ is an hermitian $N\times N$ matrix.

We are interested in computing the generating
function $\Phi(u,g)$ for the disk amplitude, defined by
\begin{equation}
  \Phi(u,g) = \sum_{k = 0}^{\infty} p_k(g) u^k
  = \sum_{k,n = 0}^{\infty} {\cal N}(k;n) g^n u^k\ ,
  \labell{eq:gf}
\end{equation}
where $p_k$ is equal to the partition function summed over all
triangulations having $k$ boundary edges. The correspondence with the
matrix model language is that ${\cal N}(k; n)$ is equal to the number
of planar diagrams (up to symmetry factors) with $n$ trivalent
vertices and $k$ external legs; these may be thought of as the dual
graphs to the triangulations. Thus, $p_k$ is
given by the matrix model expectation value
\begin{equation}
  p_k = {1\over N}\left\langle {\rm Tr}\; (U^k) \right\rangle \ .
  \labell{eq:relate}
\end{equation}
so that
\begin{equation}
\Phi(u,g)={1\over N}\left\langle{\rm Tr}\; {1\over 1-uU}\right\rangle \ .
\end{equation}
(Note: in this paper all matrix model expectation values are evaluated
in the large $N$ limit.)

There are two simple ways to solve for the function $\Phi(u,g)$. In
Ref. \cite{tutte}, a recursion relation is derived for the number of
triangulations in which no two edges join the same pair of vertices, no 
internal edge is connected to two external vertices, and no two triangles 
have three vertices in common.  From the point of view of the dual graphs,
this is equivalent to demanding that the graphs be
built from connected, one-particle irreducible diagrams
with no self-energy corrections to the propagator. 
Similar recursion relations are discussed in Ref.
\cite{bc,knn}.  Alternatively, $\Phi(u,g)$ can be evaluated by computing
the functional integral over matrices. In this case, all
Feynman graphs are summed over automatically, and the coefficients of
$\Phi(u,g)$ can be obtained by saddle point methods or by the method
of orthogonal polynomials
\cite{biz,BIPZ}. Although the numerical values of the coefficients
${\cal N}(k;n)$ are different in the Tutte and matrix model schemes, the
universal behaviour of $\Phi(u,g)$ (which governs the continuum limit) is
the same.  We shall follow Tutte in deriving recursion equations for the
number of triangulations, but we adopt the matrix model conventions for
counting: our dual graphs need not be connected nor one-particle
irreducible, and we allow self-energy corrections to the propagator.

\subsection{Generating equation}

We now proceed to derive a recursion relation for the coefficients
${\cal N}(k;n)$, which leads to a single generating equation satisfied
by $\Phi(u,g)$.  We consider a triangulation of a disk $D$ with $n$
triangles and $k$ external edges, one of which is marked.  The given
triangulation can be disassembled step by step, by repeatedly
removing the marked edge and marking a new edge on the resulting,
smaller triangulation.  (A similar approach has been used recently 
to derive a discrete Hamiltonian operator for string field theory in 
the proper time gauge \cite{kkmw,gubkleb,watabiki}.)
Each time a marked edge is removed, this has the effect of either
removing a triangle, or removing a pair of boundary edges. In this
way, for each triangulation there is a unique sequence of ``moves''
reducing to the trivial triangulation, a single dot.  By considering
all possible sequences of moves, we can recursively count the
number of triangulations ${\cal N}(k;n)$.  

\begin{figure}[t]
\centerline{
\psfig{figure=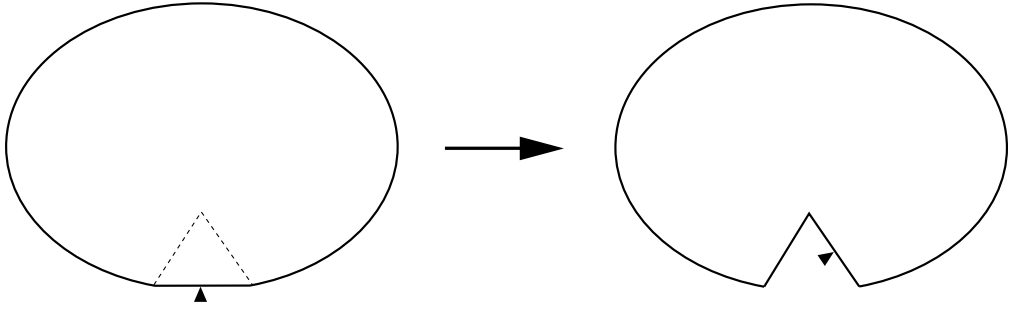,angle=0,height=4cm}}
\begin{center}
\parbox{5in}{
\caption{\em The marked edge is the boundary of a triangle. The triangle is
removed. Of the two new edges, the one further counterclockwise is
marked in the new triangulation. }\label{f:remtri} }
\end{center}
\end{figure}

\begin{figure}[t]
\centerline{
\psfig{figure=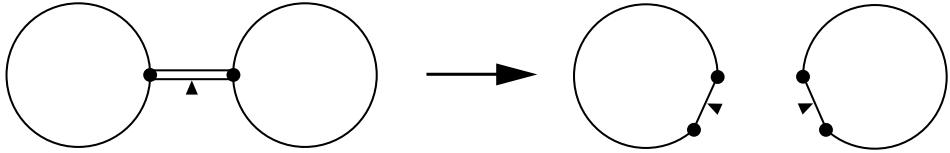,angle=0,height=2cm}}
\begin{center}
\parbox{5in}{
\caption{\em When the marked edge is connected to another boundary edge, the
triangulation is split into two disconnected pieces. Each new triangulation
has a marked edge next to where the removed edges were
attached.}\label{f:remedges}
}
\end{center}
\end{figure}

\begin{figure}
\centerline{
\psfig{figure=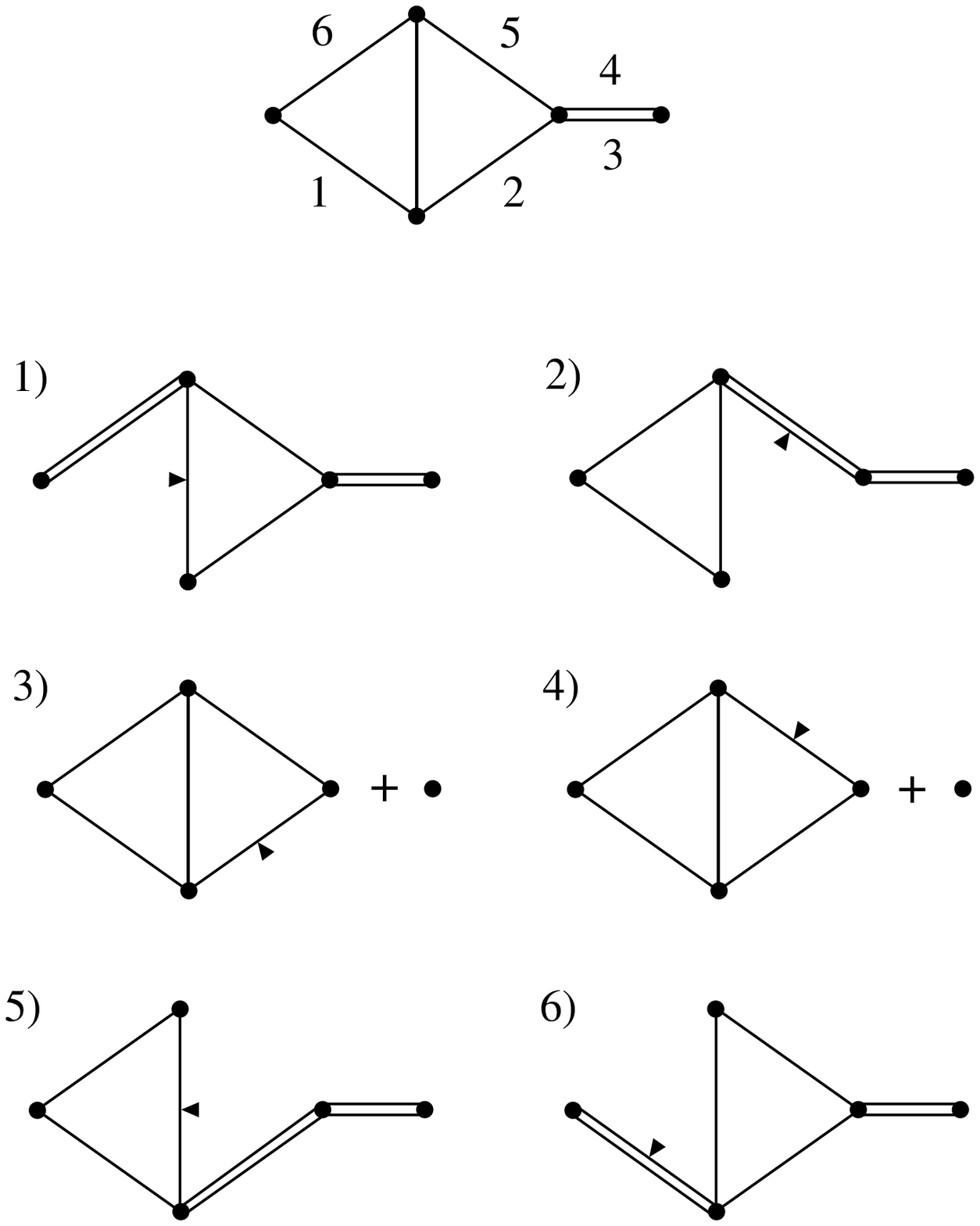,angle=0,height=8cm}}
\begin{center}
\parbox{5in}{
\caption{\em A simple triangulation is shown at top, with numbers
representing different edges which could be marked.  The six diagrams
below are the results of removing the appropriate marked edge, and
show which edge of the new triangulation becomes marked. }  \label{f:apart}
}
\end{center}
\end{figure}

Some conventions are necessary to make the recursion relations
well-defined.  When the marked edge is a boundary of a triangle,   one
removes the triangle attached to the edge; the resulting triangulation
has one less triangle, but one additional edge.  We use a convention
that marks the edge which was counterclockwise from the original marked
edge in the triangle that was removed (see Fig.
\ref{f:remtri}).  In the case where the marked boundary edge is
connected to another boundary edge, one should remove the pair of
edges, leaving two triangulations.  On each of the remaining
triangulations, we choose the convention of marking the edge 
next to edge which was just removed (see Fig.
\ref{f:remedges}).  In Fig. \ref{f:apart}, we illustrate the results
that can be obtained upon removing various possible marked edges from
a simple triangulation.

The number of triangulations with $n$ triangles and $k$ boundary
edges is therefore equal to the number with one fewer triangle and
one additional edge, plus the sum of all possible combinations of
two triangulations with the same number of triangles but a total
of two fewer boundary edges:
\begin{equation}
  {\cal N}(k;n) = {\cal N}(k+1;n-1) + \sum^{k-2}_{i=0} \sum^n_{n'=0} 
  {\cal N}(i;n'){\cal N}(k-2-i;n-n')\ .
\labell{eq:rr}
\end{equation}
In order that this recursion relation completely determine ${\cal
N}(k;n)$, it is necessary to impose boundary conditions: there is one
trivial triangulation, so that ${\cal N}(0,0)=1$, and in addition we
specify that ${\cal N}(0;n)=0$ for all $n\ge 1$.  These conditions
suffice to recover the counting of the matrix model.

Using (\ref{eq:gf}), it is straightforward to rewrite (\ref{eq:rr}) 
as an equation for the generating function $\Phi(u,g)$:
\begin{equation}
  \Phi(u,g) = 1 + {g\over u} (\Phi(u,g) - p_1(g) u -1) + u^2\Phi^2(u,g)  
  \labell{eq:ge}
\end{equation}
The first term on the right hand side of (\ref{eq:ge}) 
fixes the boundary condition ${\cal N}(0;n)=\delta_{n,0}$.
The second term corresponds to the first term in
(\ref{eq:rr}), and represents the removal of a triangle, as in
Fig. \ref{f:remtri}; the negative terms are subtracted from $\Phi(u,g)$
since no move that removes a triangle can result in a triangulation with
fewer than two edges. The third term corresponds to the second term in
(\ref{eq:rr}).

It is important to note that (\ref{eq:ge}) cannot be solved for $\Phi$
directly, since it contains the function $p_1(g)$ which is a boundary
condition for $\Phi(u,g)$ that must be computed separately. Equations of
this general form have been studied extensively in the literature, and it
is possible to show that the requirement that $\Phi(u,g)$ have a well
defined power series expansion in $u$ and $g$ around the origin is
sufficient to completely determine the power series $p_1(g)$ from
(\ref{eq:ge}). We refer the reader to Refs. \cite{tutte,bc,knn} for
details of this procedure for the present example, and for the more
complicated cases that we shall encounter in later sections.

\subsection{Graphical interpretation}

The generating equation (\ref{eq:ge}) may be written more compactly by
introducing a special derivative operator, which will turn out to be
useful in more complicated matrix models where $\Phi$ will be a
function of free variables.  We define the operation of $\partial_u$
on a polynomial in $u$ so that it removes one power of $u$ from terms
proportional to $u^k$ (without introducing a factor of $k$), and
annihilates terms which are independent of $u$
\cite{voiculescu,dl}.  Thus, 
\begin{equation}
  \partial_u \sum_{k=0}^\infty c_k u^k = \sum_{k=1}^\infty c_k u^{k-1}\ .
  \labell{eq:der}
\end{equation}
Such an operator does not obey the usual Leibniz rule; rather,
\begin{equation}
  \partial_u (\Phi \Psi) = (\partial_u \Phi) \Psi
  + \Phi(u=0) \partial_u \Psi\ .
  \labell{eq:leibniz}
\end{equation}
Using this definition, we have
\begin{equation}
  \partial_u \Phi={1\over u}(\Phi-1),\qquad \partial^2_u\Phi={1\over u^2}
  (\Phi-1-u p_1)\ ,
 \label{eq:purederivatives}
\end{equation}
and (\ref{eq:ge}) reduces to
\begin{equation}
  \Phi(u,g) = 1 + gu\partial^2_u\Phi(u,g) + u^2\Phi^2(u,g) \ .
  \labell{eq:nge}
\end{equation}

In this notation, the correspondence between the terms in the
generating equation and the graphical representation of the basic
moves becomes precise, in the sense that every symbol appearing in
eq.~(\ref{eq:nge}) has a direct interpretation.  On the right hand
side there are three terms, corresponding to a boundary condition, the
move which removes a triangle, and the move which removes two
identified boundary edges; each of these three types of terms will
appear in the more elaborate generating equations we present in the
remainder of the paper.  The right hand side of each such equation
will begin with a one, which sets the number of triangulations with no
boundary edges equal to one.  The terms corresponding to removal of a
triangle will have a factor of the weight for a single triangle (in
this case, $g$) and an operation in free variables (in this case,
$u\partial^2_u$) which represents replacing the single marked edge
with two new edges.\footnote{The notation $u\partial^2_u$ may seem to suggest
replacing two edges with a single one, rather than the other way around.
This is merely because we have been using language in which the
triangulations represented by $\Phi$ are disassembled step by step, rather
than being built up from nothing; but either sense is equally legitimate.}
  The terms corresponding to removing two
identified boundary edges will contain a variable for each boundary
edge (in this case, two factors of $u$) and two functions (in this
case, two factors of $\Phi$) representing the two triangulations that
remain after the removal of the boundary edges.  In more complicated
models these terms may be multiplied by coupling constants
representing the interaction between matter fields, as we shall see in
the next section.

Eq. (\ref{eq:nge}) is closely related to the Schwinger-Dyson equation for
the 1-matrix model, which can be written as
\begin{equation}
  \partial_u\Phi = u\Phi^2 + g \partial_u^2\Phi\ .
  \labell{eq:1msd}
\end{equation}
It is easy to see that acting by $\partial_u$ on (\ref{eq:nge}) yields
(\ref{eq:1msd}). The generating equation includes the information contained
within the Schwinger-Dyson equation, as well as additional boundary
conditions.  This will continue to be the case for more complicated
theories described by multi-matrix models, where we shall derive a single
equation equivalent to the entire set of Schwinger-Dyson equations plus
boundary conditions.

\subsection{Higher genus}

It is reasonably straightforward to extend the recursion relation
(\ref{eq:rr}) from the disk amplitude to an amplitude for a higher
genus surface with an arbitrary number of boundaries.  We imagine that
the set of boundaries is ordered, and that each one has a marked edge,
and derive a recursion equation by removing the marked edge on the
first boundary loop.  The number of triangulations of a surface with
genus $G$ and with $b$ boundaries, represented by
\begin{equation}
  {\cal N}^{(b)}_G(k_1,k_2,\ldots,k_b;n)\ ,
\end{equation}
with all $k_a>0$, is given by the equation
\begin{equation}
  \begin{array}{l}
  \displaystyle{{\cal N}^{(b)}_G(k_1,k_2,\ldots,k_b;n)\ = 
  \ {\cal N}^{(b)}_G(k_1+1,k_2,\ldots,k_b;n-1)}
  \\
  \displaystyle{\quad + \sum_{a=0}^{b-1} \sum_{\{\sigma_i\}}
  \sum_{H=0}^G \sum_{i=0}^{k_1-2} \sum_{m=0}^n
  {\cal N}^{(a+1)}_H(i,k_{\sigma_1},\ldots,k_{\sigma_a};m) 
  {\cal N}^{(b-a)}_{G-H}(k_1-i-2,k_{\sigma_{a+1}},\ldots,k_{\sigma_{b-1}};n-m)}
  \\
  \displaystyle{\quad + \sum_{a=2}^b k_a
  {\cal N}^{(b-1)}_G(k_1+k_a-2,k_2,\ldots,k_{a -1},k_{a + 1},\ldots,k_b;n-1)}
  \\
  \displaystyle{\quad + \sum_{i=0}^{k_1-2} 
  {\cal N}^{(b+1)}_{G-1}(i,k_1-i-2,k_2,\ldots,k_b;n)}
  \end{array}
  \labell{highergenusrec}
\end{equation}
where $\{\sigma_i\}$ in the second term represents the set of all
permutations of $(2,3, \ldots, b)$ such that $\sigma_1<\sigma_2<\ldots
<\sigma_a$ and $\sigma_{a+1}< \sigma_{a+2}<\ldots <\sigma_{b -1}$, and
we set ${\cal N}^{(b)}_G(k_1,k_2,\ldots,k_b;n)=0$ when any of the
$k_a$'s are zero (except that, as before, ${\cal N}^{(1)}_0(0;0)=0$).

\begin{figure}
\centerline{
\psfig{figure=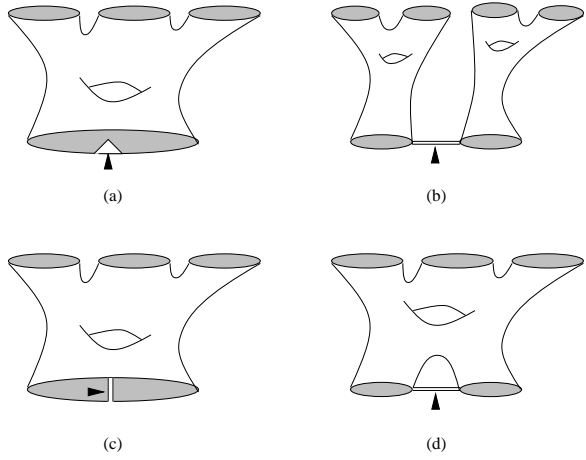,angle=-90,height=6cm}}
\begin{center}
\parbox{5in}{
\caption{\em For a higher genus surface, there are two new moves (c) and (d),
as well as the two moves (a) and (b) needed for the disk amplitude.}
\label{f:topology}
}
\end{center}
\end{figure}

Once again, each term in this expression has a geometric interpretation
(see Fig \ref{f:topology}).  The first term on the right hand side
corresponds to the removal of a single triangle (Fig \ref{f:topology}a),
while the remaining three correspond to the removal of an identified pair
of boundary edges.  The second term is related to the second term in
(\ref{eq:rr}), derived from the splitting of a single triangulation into
all possible combinations of two (Fig \ref{f:topology}b).  The third term
corresponds to the removal of two identified edges which belong to
different boundary components; the result is to decrease the number of
boundaries by one (Fig \ref{f:topology}c).  The last term comes from the
removal of two identified edges belonging to the same boundary component,
such that the resulting triangulation does not split into two; the result
is to increase the number of boundaries by one, while decreasing the genus
by one (Fig \ref{f:topology}d).

One can define generating functions
\begin{equation}
  \Phi^{(b)}(u_1,u_2, \ldots, u_b,g,h)= \sum_{G=0}^\infty 
  \sum_{\{k_j\},n=0}^\infty
  h^G{\cal N}^{(b)}_G(k_1,k_2, \ldots, k_b;n)\;g^n u_1^{k_1}
  u_2^{k_2}u_3^{k_3}\cdots 
  u_b^{k_b} 
\end{equation}
for triangulations of a surface with $b$ boundaries of arbitrary genus.
It is now possible to convert the recurrence relation
(\ref{highergenusrec}) into an equation that relates amplitudes for
surfaces of arbitrary genus with different numbers of boundaries:
\begin{eqnarray}
\lefteqn{  \Phi^{(b)}(u_1, \ldots, u_b,g,h)=}\nonumber \\
&& \delta_{b1} + g
  u_1\partial_{u_1}^2\Phi^{(b)}(u_1, \ldots, u_b,g,h)  
  \nonumber  \\
  &&
  + u_1^2\sum_{a=0}^{b-1}\sum_{\{\sigma_i\}}
  \Phi^{(a+1)}(u_1,u_{\sigma_1}, \ldots,
  u_{\sigma_{a}},g,h)\Phi^{(b-a)}(u_1,u_{\sigma_{a+1}}, \ldots, u_{b},g,h) 
  \nonumber  \\
  && + 
   \sum_{j=2}^bu_j{d\over du_j} \left[ u_1 u_j
D_{u_1,u_j}\Phi^{(b-1)}
  (u_1, \ldots, u_{j-1},u_{j+1}, \ldots, u_b,g,h)  \right]
 \labell{eq:geg}  \\
  && +    h u_1^2 \sum_{j=2}^b
  \Phi^{(b+1)}(u_1, \ldots, u_{j-1},u_1,u_{j}, \ldots, u_{b},g,h)\ ,
\nonumber
\end{eqnarray}
where $\{\sigma_i\}$ is defined as before,
the operator $D_{u_1,u_j}$ is defined by
\begin{equation}
  \begin{array}{l}
  D_{u_1,u_j}\Phi^{(b-1)}\left(u_1, \ldots, u_{j-1},u_{j+1}, \ldots, u_b,g,h
  \right) = 
  \\
  \displaystyle
  {\qquad{ u_1\Phi^{(b-1)}(u_1, \ldots, u_{j-1},u_{j+1}, \ldots, u_b,g,h) -
  u_j\Phi^{(b-1)}(u_2, \ldots, u_b,g,h)\over u_1-u_j}}\ ,
  \end{array}
\end{equation}
and $d/du_i$ is an ordinary derivative operator.

Equation (\ref{eq:geg}) is useful in part because it contains a set of
linear relations that permit an iterative solution for generating functions
of arbitrary genus and with an arbitrary number of surfaces starting from
the solution of Eq. (\ref{eq:ge}) for the disk. For example, the simplest
relation contained in that equation is obtained by projecting out the genus
zero component $h^0$ and taking $b=2$. This gives a relation between the
disk and cylinder generating functions (the $h^0$ components $\Phi^{(1)}_0$
and $\Phi^{(2)}_0$ of $\Phi^{(1)}$ and $\Phi^{(2)}$):
\begin{equation}
\begin{array}{rcl}
  \Phi^{(2)}_0(u_1,u_2,g) 
  &=& 2 u_1^2
  \Phi^{(1)}_0(u_1,g)\Phi^{(2)}_0(u_1,u_2,g) + gu_1\partial_{u_1}^2
  \Phi^{(2)}_0(u_1,u_2,g) + 
  \\ 
  && \displaystyle \quad u_2{d\over
  du_2}\left({u_1^2 u_2\Phi^{(1)}_0(u_1,g)-u_2^2 u_1\Phi^{(1)}_0(u_2,g)\over
  u_1-u_2}\right)\ .
\end{array}
\end{equation}
Another simple relation exists between the generating functions for the
disk, the cylinder and the single-boundary torus (the $h^1$ component
taking $b=1$):
\begin{equation}
  \Phi^{(1)}_1(u_1,g) = 2 u^2 \Phi^{(1)}_0(u_1,g)\Phi^{(1)}_1(u_1,g)
  + g u_1\partial_{u_1}^2\Phi^{(1)}_1(u_1,g) 
  + u_1^2 \Phi^{(2)}_0(u_1,u_1,g)\ .
\end{equation}
This procedure may be continued to obtain equations for arbitrary
genus and number of boundary components.

\section{Ising matter}
\label{sec:ising}

We now consider 2D gravity coupled to the Ising model.  This model
corresponds to the $c = 1/2$ minimal model CFT coupled to Liouville
gravity, and has a simple description as a two-matrix model.  As in
the previous section, most of the results we describe here have long
been understood in terms of the matrix model description of the
theory; the novelty of our presentation lies in the purely geometric
derivation of a single generating equation written in terms of
noncommuting variables.

\subsection{Review of model}

The coupling of the Ising model to gravity is described by summing over all
triangulations of a closed 2-manifold just as in the pure gravity model.
Now, however, each triangle carries a label $U$ or $V$, corresponding to
Ising spins up or down respectively.  The weight of a given triangulation
$\Delta$ is defined to be 
\begin{equation}
  W (\Delta) =
  \frac{1}{S (\Delta)}  N^{\chi (\Delta)} g^{n (\Delta)}
  \left( \frac{1}{1-c^2} \right)^{uu (\Delta)+ vv (\Delta)}
  \left( \frac{c}{1-c^2} \right)^{uv (\Delta)}
\end{equation}
where the notation is the same as in (\ref{eq:pgpf}), with the
addition of a new coupling constant $c$;
$uu (\Delta)$ is the number of edges in $\Delta$ separating
two $U$ triangles, and similarly for $uv(\Delta)$ and $vv(\Delta)$.
The coupling $c$ is related to the usual Ising model coupling $J$ by
\begin{equation}
  c = e^{-2J}\ .
  \labell{eq:candj}
\end{equation}
We use $c$ rather than $J$ because this is the natural notation in
the description of the theory as a 2-matrix model, with partition
function given by \cite{kazakov,boukaz}
\begin{equation}
Z = \sum_{\Delta} W (\Delta)
= \int {\rm D} U \; {\rm D} V \;
\exp \left(-NS (U,V) \right)\ ,
\labell{eq:matrixising}
\end{equation}
with
\begin{equation}
S (U,V)=\frac{1}{2} {\rm Tr}\; U^2 + 
\frac{1}{2}{\rm Tr}\;  V^2 -c{\rm Tr}\; UV -
\frac{g}{3}  ({\rm Tr}\; U^3 +{\rm Tr}\; V^3)\ ,
\label{eq:}
\end{equation}
where $U$ and $V$ are $N\times N$ hermitian matrices.

\subsection{Generating equation}

\begin{figure}
\centerline{
\psfig{figure=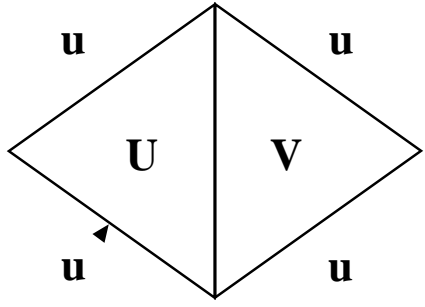,angle=0,height=4cm}}
\begin{center}
\parbox{5in}{
\caption{\em A diagram contributing 
a factor of $N g^2 c^3/(1-c^2)^5$ to the expectation value 
$\langle {\rm Tr}\; U^4\rangle = N p_{uuuu}$.}\label{f:term}
}
\end{center}
\end{figure}

We would like to analyze the generating function associated with a disk, as
in the pure gravity theory discussed in Section 2.  However, in this model
it is necessary to specify the configuration of Ising spins on the boundary
rather than just the number of edges.  Thus, we must associate with each
boundary a length $l$, and a string of $l$ labels, each of which is either
a $u$ or a $v$.  These labels may be thought of as spins lying {\em
outside} the boundary of the triangulation, meaning that each boundary edge
contributes a weight $1/(1-c^2)$ or $c/(1-c^2)$ depending on whether the
spin on the boundary edge is the same as or different from the spin on the
triangle adjacent to that edge; although this may seem to be a
counterintuitive convention, it will simplify the following analysis.  By
choosing an orientation and a marked edge on the boundary, we have a
natural ordering to the boundary labels; this ordering is given by
beginning with the marked edge and proceeding around the boundary.  It is
now natural to consider $u$ and $v$ as noncommuting variables; the
generating function for the disk can then be written as (throughout this
section we suppress the dependence of $\Phi$ on the couplings $g,c$ and the
noncommuting variables $u,v$)
\begin{equation}
\begin{array}{rcl}
  \Phi &=& 1 +p_u (u + v) + p_{uu}(u^2 + v^2)
  +p_{uv}(uv+vu) \\
  && + p_{uuu}(u^3 + v^3)
  +p_{uuv}(uuv+uvu+vuu+uvv+vuv+vvu) + \ldots
\end{array}
\labell{eq:pees}
\end{equation}
The coefficients $p_{w(u,v)}$ represent the disk amplitude
for triangulations with boundary condition specified by 
$w(u,v)$, a word in the free algebra generated by $u$ and $v$
({\em i.e.}, an ordered product of $u$'s and $v$'s).  Notice that, due
to the symmetry between plus and minus spins, the expectation values
are related by $p_{w(u,v)} = p_{w(v,u)}$.  (In the models we shall
study in subsequent sections, this will no longer be the case.)
There are also symmetries under cyclic permutation and reversal of
$w(u,v)$.  In matrix model language, $p_{w(u,v)}$ is just 
the usual expectation value of a trace of a product of hermitian
matrices (in the large $N$ limit):
\begin{equation}
  p_{w(u,v)} = {1\over N}\left\langle {\rm Tr}\; w(U,V) \right\rangle \ ,
  \labell{eq:relate1}
\end{equation}
and $\Phi$ is the generating function of planar
Green's functions \cite{cvitanovic,gg,dl}:
\begin{equation}
\begin{array}{rcl}
  \Phi & = & \displaystyle\frac{1}{N}
  \sum_{n = 0}^{\infty}  \left\langle {\rm Tr}\; (uU + vV)^n
  \right\rangle\\
  &=& \displaystyle\frac{1}{N}\sum_{w (u,v)}
  w (u,v) \left\langle {\rm Tr}\; w (U,V) \right\rangle\ .
\end{array}
\end{equation}
For example, Figure~\ref{f:term} illustrates a diagram contributing 
a factor of $N g^2 c^3/(1-c^2)^5$ to the expectation value 
$\langle {\rm Tr}\; U^4\rangle = N p_{uuuu}$. 

We can now, as in the previous section, use purely geometrical arguments 
to derive a single generating equation satisfied by $\Phi$, 
taking advantage of the interpretation discussed after 
Eq.~(\ref{eq:nge}) to write down the generating equation directly,
without first deriving a recursion relation.  We have
associated the first symbol in a word $w (u,v)$ with a marked edge on the
boundary of a triangulation.  If we remove this edge, then as in the pure
gravity case there are two possibilities: either the edge is identified
with another edge of the boundary (which must also be removed), or the edge
is attached to a triangle which can be removed.  Unlike the pure gravity
case, however, in each of these cases there are multiple possible spin
configurations which must be included.  

\begin{figure}
\centerline{
\psfig{figure=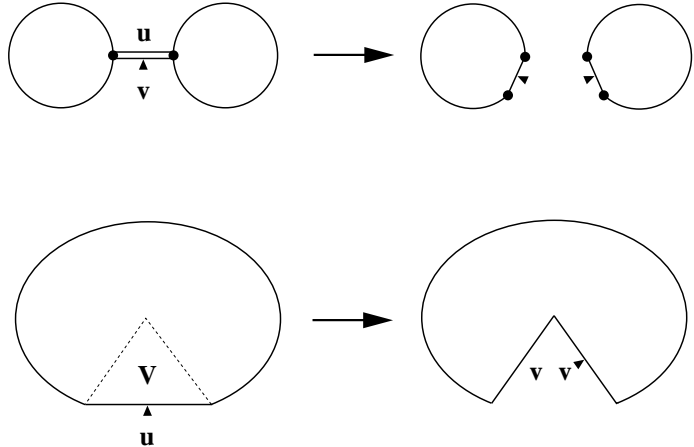,angle=0,height=6cm}}
\begin{center}
\parbox{5in}{
\caption{\em The removal of a pair of edges of opposite spin results in a
term $cv\Phi u\Phi$ on the right hand side of the generating equation. The
removal of a triangle with a spin at its center not equal to the spin on
its edge results in a term $cgu\partial_v^2\Phi$. Note that the two new
edges carry the spin of the triangle that was removed.}\label{f:diagram48}
}
\end{center}
\end{figure}

It is useful to define derivative operators $\partial_u$ and $\partial_v$
that are similar to the derivative operator defined in the previous
section.  (Operators of this form were introduced in Refs. 
\cite{voiculescu,dl}.)
If $\partial_u$ ($\partial_v$) acts on a word with a $u$ ($v$) at the
leftmost position, that $u$ ($v$) is removed; if a $v$ ($u$) is at the
leftmost then the term is annihilated. For example:
\begin{equation}
\partial_u (uuuvv + vvu) = uuvv, \qquad \partial_v (uuuvv +vvu) = vu\ .
\end{equation}
In terms of these operators, the  generating
equation is
\begin{equation}
\Phi = 1 +
\frac{1}{1-c^2} \left[ u \Phi u \Phi + v \Phi v \Phi + cu \Phi v \Phi
+ cv \Phi u \Phi 
+ gu \partial_u^2 \Phi + gv \partial_v^2 \Phi + cgv \partial_u^2 \Phi
+ cgu \partial_v^2 \Phi \right].
\labell{eq:generateising}
\end{equation}
Again, each term corresponds to a possible outcome of removing a
boundary edge.  The first four terms in the square brackets are associated
with the removal of two identified edges: a marked $u$ 
identified with another $u$, a marked $v$ identified with another $v$,
etc.
(Note the factor of $c$ appearing when a $u$ is identified with
a $v$.)  The other four terms correspond to removing a triangle
(and therefore have a factor of $g$); they represent the four possibilities
of a $u$ or $v$ marked edge being identified with a $u$ or $v$ triangle.
A diagrammatic description of the fourth and eighth terms inside the
brackets is given in Figure~\ref{f:diagram48}.

The generating equation (\ref{eq:generateising}) is closely related to
the Schwinger-Dyson equations for the  corresponding matrix model.
The Schwinger-Dyson equations are given by  \cite{dl}
\begin{equation}\begin{array}{rcl}
\partial_u \Phi & = &  \Phi u \Phi + g \partial_u^2 \Phi +c \partial_v \Phi\\
\partial_v \Phi & = & \Phi v \Phi + g \partial_v^2 \Phi + c \partial_u \Phi.
\end{array}
\end{equation}
It is fairly easy to check that these equations are equivalent to
(\ref{eq:generateising}) except that they do not determine the leading
constant in $\Phi$.  Thus, the single generating equation contains all the
information contained in the pair of Schwinger-Dyson equations.  Having a
single equation for $\Phi$ in terms of noncommuting variables simplifies
the subsequent analysis of such a system, as we shall see in the following
sections.  Finally, note that it is straightforward to write down the
generalization of the higher genus generating equation (\ref{eq:geg}) in
the presence of Ising matter. 

\subsection{Solving the generating equation}

Now that we have a single generating equation describing the set of all
disk amplitudes, we would like to find a closed form solution for $\Phi$ in
terms of noncommuting variables.  Unfortunately, however, finding a
complete solution to such a nonlinear equation in noncommuting variables
seems to be a  difficult problem \cite{gg,dl}.  It is possible to
directly compute any specific disk amplitude with fixed boundary length $l$
and a finite power of $g$ by repeated application of
(\ref{eq:generateising}).  However, to understand the continuum limit of
the theory we need to have information about amplitudes with arbitrarily
large boundaries.  In certain cases it has been found possible to find a
closed set of loop equations for matrix models which allow the computation
of generating functions with certain constraints on boundary configurations
\cite{gn,alfaro,staudacher}.  We shall now demonstrate in the generating
function approach a simple systematic procedure by which such closed
systems of equations can be found and solved.  The following analysis
closely parallels that of \cite{dl}.

By factoring out leading terms which depend on $v$,
the generating function $\Phi$ can be expanded in terms of functions
$\phi_{w(u,v)}$ which only depend on $u$:
\begin{equation}\begin{array}{rcl}
  \Phi (u,v) &=& \phi (u) + v \phi_v (u) + v^2 \phi_{vv} (u) +
  uv \partial_u \phi_v (u) \\
  && + v^3 \phi_{vvv}(u) + vuv \phi_{vuv}(u) 
  + uvv \partial_u\phi_{vv}(u) + uuv \partial_u^2\phi_v(u) + \cdots
\end{array}
\labell{eq:pexpand}
\end{equation}
The functions $\phi_{w(u,v)}(u)$ are defined by
\begin{equation}
  \phi_{w(u,v)} = \left[w(\partial_u,\partial_v) \Phi\right]_{v=0}\ ,
  \labell{eq:extraction}
\end{equation}
where $w(u,v)$ begins and ends with $v$.  (If $w(u,v)$ begins with
a $u$, we can use the cyclic symmetry of the generating function to
replace the corresponding function $\phi_{w(u,v)}$ by
a derivative of a lower-order function, as we have done in 
(\ref{eq:pexpand}).)  Thus, $w(u,v)\phi_{w(u,v)}$ is that piece of
$\Phi$ containing words consisting of $w(u,v)$ multiplied on the
right by a string of $u$'s.  (Note that (\ref{eq:extraction}) actually
extracts the term with the $u$'s and $v$'s in the opposite order, but
the functions are equal by symmetry.)

Considering only those terms on each side of
(\ref{eq:generateising}) with a fixed initial string followed by
$u$'s, we can write a closed system of equations for 
a subset of the functions $\phi_{w(u,v)}(u)$.  
For example, taking only terms with $u$'s, we have
\begin{equation}
  \phi = 1+{1\over{1-c^2}}\left( u^2\phi^2 +gu\partial_u^2\phi 
  + cgu \phi_{vv}\right)\ .
  \labell{eq:firstphi}
\end{equation}
Taking terms of the form $v u^i$, $uvu^j$, $v^2 u^k$ give respectively
\begin{eqnarray}
  \phi_v & = & {1\over{1-c^2}}\left(cu \phi^2 + g\phi_{vv} 
  + cg \partial_u^2 \phi\right)
  \nonumber \\
  \partial_u\phi_v
   & = & {1\over{1-c^2}}\left(u \phi \phi_v + c \phi 
  + g \partial_u^2 \phi_v + cg \phi_{vvv}\right) 
  \labell{eq:firstphi2}   \\
  \phi_{vv}  & = & {1\over{1-c^2}}\left(\phi + cu \phi_v \phi + 
  g \phi_{vvv} + cg \partial_u^2 \phi_v\right)\ .
  \nonumber
\end{eqnarray}
This gives us 4 equations in 4 unknowns, each of which is a function
of the single commuting variable $u$.  
These 4 equations can be combined to give a single equation  for
$\phi$ and its derivatives:
\begin{equation}\begin{array}{rcl}
  0 & = & gu^3 \phi^3+ (cu^2- g u+ g^2) \phi^2  -g^2
  + (c^3u-cu-gu^2 \phi) \partial_u \phi\\
  & & + (gu + cgu + g^2 u^2 \phi) \partial_u^2 \phi
  -2 g^2u \partial_u^3\phi+ g^3u \partial_u^4\phi\ .
\end{array}\labell{eq:cubic1}\end{equation}
As was done in Eq.~(\ref{eq:purederivatives}) for pure gravity, we can write
the derivatives of $\phi$ in terms of $\phi$, $u$, and a finite
number of the individual coefficients $p_u$, $p_{uu}$, and so on.
This allows us to rewrite (\ref{eq:cubic1}) as a pure cubic in $\phi$:
\begin{equation}
  f_3 \phi^3 + f_2 \phi^2 +f_1 \phi + f_0 = 0\ ,
  \labell{eq:purecubic}
\end{equation}
where
\begin{equation}
  \begin{array}{rcl}
  f_0 &=&   -{g^3}+g^2 \left( 2 - {g}p_1 \right) u  -
   g\left( 1 + c - 2{g}p_1 + {g^2}p_2 \right) 
    {u^2} 
  \\
  &&\qquad -\left( {g^2} + g(1+c)p_1 - 
      2{g^2}p_2 + {g^3}p_3 -c(1-{c^2})\right) {u^3}
  \\
  f_1 &=&  {g^3} -2{g^2}u +g\left(1+c \right) {u^2}  -
   \left( c(1 - {c^2}) + {g^2} \right) {u^3} +
   g\left(1- {g}p_1 \right) {u^4}
  \\
  f_2 &=& u^3(2 g^2 -2 g u + c  u^2)
  \\
  f_3 &=& g u^6
  \end{array}
\end{equation}
Notice that the coefficients $p_u$ and $p_{uu}$ appearing here are not
independent; the order $u$ terms in (\ref{eq:firstphi}) give
\begin{equation}
  p_{uu}={(1-c)\over g} p_u \ .
\end{equation}

The cubic (\ref{eq:purecubic}) describing homogeneous (fixed spin) disk
amplitudes for the Ising model coupled to gravity is well known and
has been derived using a variety of matrix model methods
\cite{gn,alfaro,staudacher,dl}.  The derivation we have given is completely
independent of the matrix model formalism, and depends only on the
underlying geometric theory.  The existence of a single equation for
the noncommuting generating function (\ref{eq:generateising}) rather
than a pair of Schwinger-Dyson equations has simplified our analysis.

It is straightforward given the disk amplitude to obtain expressions for
generating functions in commuting variables for boundary data with multiple
domains of like spins. One proceeds iteratively, deriving an expression for
the two domain amplitude in terms of the homogeneous amplitudes and so on
\cite{staudacher}. Defining $\tau(u,v)$ as the two-domain generating
function, including all terms homogeneous in $u$ and $v$, it follows from
(\ref{eq:generateising}) that
\begin{eqnarray}
  \tau-\phi(v)&=&{1\over1-c^2}\Bigl[u\phi(u)u\tau(u,v)
  +cu\tau(u,v)v\phi(v) \nonumber \\
  &&\quad +{g\over u}(\tau(u,v)-u\phi_v(v)-\phi(v)) + {{cgu}\over{v^2}}
  (\tau(u,v)-v\phi_v(u) -\phi(u))\Bigr]
\end{eqnarray}
which is linear in $\tau(u,v)$.  (Note that in this equation $u$ and $v$ act
as ordinary commuting variables, and we have converted the derivative 
operators into their algebraic expressions.) Similar relations can easily 
be derived for the higher domain generating functions.

\subsection{Alternate boundary conditions}
\label{sec:boundarycorrelation}

We have shown how noncommuting variables can be used to encode the
boundary data used in a sum over triangulations coupled to Ising
matter.  However, the decision to describe the boundary as a sequence
of plus and minus spins is not unique.  In this subsection we briefly
describe an alternative description of the Ising theory, in which the
boundary condition is specified by labelling the vertices between
boundary edges as either ``stick'' or ``flip,'' depending on whether
the spins on either side are like or unlike.  We outline  the
derivation of
the cubic equation corresponding to all-stick boundary conditions,
which is equivalent to (\ref{eq:purecubic}).  Although we do not
derive any results here which cannot be found using standard matrix
model methods, there are certain correlation functions which can be
derived using this alternative formulation which appear to be
inaccessible using the matrix model language.  An example of such a
calculation appears in \cite{cot2}.

The generating equation in the flip/stick variables may be derived
using the kind of geometric argument used to find
(\ref{eq:generateising}).  For example, using the symbols $f$ and $s$
for flip and stick, the term $cgf^2\partial_s^3\Phi/(1-c^2)$
will arise when a marked edge bounded on each
side by a flip is connected to a triangle, which when removed reveals
two edges (and thus three vertices) labelled by $sss$.  This
corresponds in the $u$, $v$ variables to removing a plus spin which is
surrounded by two minuses and connected to a minus triangle, or the
same operation with plus and minus interchanged.  Taking all possible
such moves into account, we obtain
\begin{eqnarray}
\lefteqn{\Phi = 1 +{1\over 1-c^2}
  \Biggl\{g[(s+ c f^2\partial_s)+(c f s + s f)\partial_f]
  \partial^2_s\Phi + g[(f s + c s f)\partial_s+(c s^2+f^2)\partial_f]
  \partial_s\partial_f\Phi}\nonumber \\
  & &+{1\over 2}
  \left[(c s^2+f^2)(\partial_s\Phi f + \partial_f\Phi s)+
  (f s+ c s f)(2+ \partial_s\Phi s + \partial_f\Phi f)\right]
  (f \partial_s\Phi + s \partial_f\Phi) \nonumber \\
  & &+  {1\over 2}\displaystyle\left[(s^2+ c f^2)
  (2+\partial_s\Phi s + \partial_f\Phi f)+
  (c f s + s f)(\partial_s\Phi f + \partial_f\Phi s)\right]
  (2+s \partial_s\Phi + f \partial_f\Phi) \nonumber\\
  & &+g(1+c)s p_{ss}\Biggr\} \labell{eq:generatefs}
\end{eqnarray}
The leading term of $\Phi$ is 1,
but all other terms appear twice corresponding to the symmetry induced by
interchanging pluses and minuses on the boundary.

{}From here, we can define as before a set of functions $\phi_{w(s,f)}(s)$
and look for a system of equations which close.  One such set of four
equations is given by
\begin{eqnarray}
  \phi_- & = & 1 + {1\over 1-c^2}\left[
  {1\over 2}s^2\left(2+s\partial_s\phi_-\right)^2 
  +g\left(s^2\partial_s^3\phi_- + s p_{ss}\right)
  +cgs\left(s\phi_{fsf}+p_{ss}\right)\right] \nonumber\\
  \phi_{ff} & = & {1\over 1-c^2}\left[{1\over 2}c
  \left(2 + s\partial_s\phi_-\right)^2 + 
  g\phi_{fsf}+cg\partial_s^3\phi_- \right]\nonumber\\
  \phi_{ff} & = & {1\over 1-c^2}\left[{1\over 2}s
  \left(2 + s\partial_s\phi_-\right)\left(s\phi_{ff}
  +p_s\right)\right.\labell{eq:fsp}\\ 
  & &\qquad \left.
  +c\left(2+s\partial_s\phi_-\right)+ cg\left(s\phi_{fssf}+p_{sss}\right)
  +g\left(s\partial^2_s\phi_{ff}+p_{fsf}\right)\right]\nonumber\\
  \phi_{fsf} & = & {1\over 1-c^2}\left[\partial_s\phi_- +
  {1\over 2}c\left(2+s\partial_s\phi_-\right)
  \left(p_s+s\phi_{ff}\right)\right.\nonumber\\
  & &\qquad \left.+g\phi_{fssf} + cg\partial_s^2\phi_{ff}\right]\nonumber
\end{eqnarray}
where the homogeneous disk amplitude with all sticks is represented by
$\phi_-(s)$.  These equations are exactly equivalent to the four equations
in (\ref{eq:firstphi}) and (\ref{eq:firstphi2}) as can be seen by noting
the relations
\begin{equation}
\begin{array}{rcl}
&& \phi_-=2\phi-1,\quad
s\phi_{ff}+p_{s}=2\phi_v,\quad s\phi_{fsf}+p_{ss}=2\phi_{vv},\quad
s\phi_{fssf} + p_{sss} =2\phi_{vvv}\\
&& p_{s}=2p_{v},\quad p_{ss}=2 p_{vv},\quad p_{sss}=2 p_{vvv}
\end{array}
\end{equation}
where the factors of 2 arise because of the interchange symmetry. The cubic
equation obtained by solving (\ref{eq:fsp}) for $\phi_-(s)$ is therefore
exactly equivalent to the equation for the homogeneous disk amplitude
$\phi(u)$.

Again, it is straightforward to extend the generating equation
(\ref{eq:generatefs}) to arbitrary genus. This extension turns out to be
useful for computing the correlation functions of operators on the sphere
\cite{cot2}.

\section{The dual Ising theory}
\label{sec:dual}

One of the most remarkable properties of the Ising model on a fixed
lattice is the
Kramers-Wannier duality symmetry, which states that the partition
function of the Ising model on a lattice $L$ at a fixed temperature
$T$ is equal to the partition function on the dual lattice $\hat{L}$
at a temperature $T' \sim 1/T$.  By putting spins on the {\em
vertices} of a random triangulation, rather than on the triangles, it
is possible to construct a dual version of the
random-surface Ising theory
discussed in the previous section.  The resulting model can be
written as a two-matrix model, and is a special case of a more general
O$(n)$ loop model \cite{kostov1,gk,duk,ks,ez,johnston}.  
In this section we  consider this
dual model, and show that the homogeneous disk amplitude of the dual
model satisfies a cubic equation very similar to  (\ref{eq:cubic1});
an equivalent cubic was derived in \cite{ez} using different methods. 
This homogeneous disk amplitude for the dual model is
precisely equal to the disk amplitude of the original Ising 
theory with free boundary conditions.

\begin{figure}
\centerline{
\psfig{figure=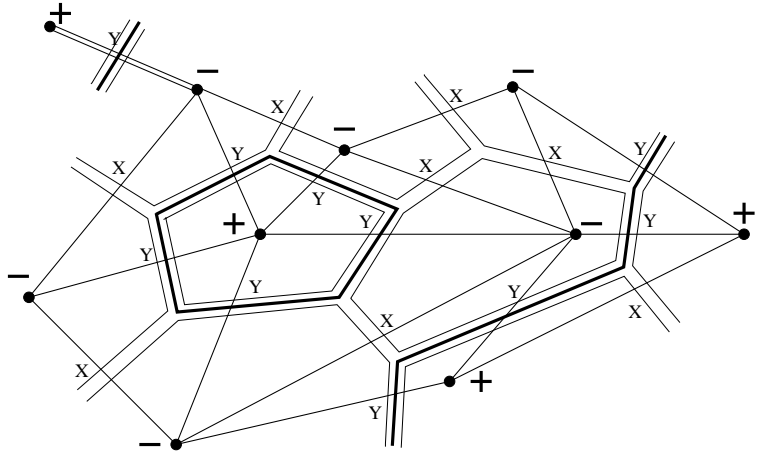,angle=-90,height=6cm}}
\begin{center}
\parbox{5in}{
\caption{\em A sample triangulation contributing to the dual model. The
dual graph, consisting of polygons which meet at trivalent vertices, 
is shown with doubled lines.  Putting the spins on the dual graph rather
than the original triangulation results in a different coupling between
the matter and gravity in the two models.
The edges labeled $Y$ (denoted by a thick line inside the
double lines) define the boundaries between
domains of like spin, and thus determine the matter configuration up 
to a sign.  Note that a boundary condition with all $X$'s
in the dual theory is equivalent to a homogeneous boundary configuration
for the dual spins.} \label{f:dualgraph}
 }
\end{center}
\end{figure}

\subsection{Dual model}

In the dual model, we associate a spin with each polygon in the dual
triangulation.  The model can be described in terms of combinatorial
data on the original triangulation by taking advantage of the fact
that any dual spin configuration is determined (up to an overall sign)
by a loop graph on the original triangulation -- that is, by a set of
edges which divide regions of opposite spin (see
Figure~\ref{f:dualgraph}).  Each edge of the triangulation (or the
corresponding edge of the dual triangulation) is labeled with a $Y$ if
it separates different spins ({\it i.e.,} if a domain boundary passes
through it), and with an $X$ if it connects identical spins.  We then
sum over all loop graphs on triangulations by summing over all
triangulations and labelings such that the boundary edges of each
triangle either have 3 $X$ labels, or 1 $X$ label and 2 $Y$ labels.
The weight given to each triangulation is
\begin{equation}
W (\Delta) =
\frac{1}{S (\Delta)}  N^{\chi (\Delta)} \hat{g}^{n (\Delta)}
\left( \frac{1}{1-c} \right)^{x (\Delta)}
\left( \frac{1}{1 +c} \right)^{y (\Delta)}\ ,
\labell{eq:dualpart}
\end{equation}
where $x (\Delta)$, $y (\Delta)$ are the number of $X,Y$ edges in
$\Delta$ respectively.  Putting the spins on the vertices of the
triangulation leads to a different coupling of the Ising model to
gravity than that considered in the previous section (see Figure
\ref{f:dualgraph}). We shall continue to refer to the model previously
introduced as the Ising theory, and we will refer to the theory
defined by Eq.~(\ref{eq:dualpart}) as the dual theory.

A matrix model description of the partition function of  the dual
theory is \cite{duk,ez}                         
\begin{equation}
  Z = \int {\rm D} X \; {\rm D} Y \;
  \exp \left(-N\left[ \frac{1-c}{2} {\rm Tr}\; X^2 + \frac{(1 + c)}{2}
  {\rm Tr}\; Y^2 
  -\frac{\hat{g}}{3} ({\rm Tr}\; X^3 + 3 {\rm Tr}\; XY^2)
  \right] \right)\ .
\labell{eq:matrixdual}
\end{equation}
It is a remarkable fact that the matrix model (\ref{eq:matrixdual}) is
precisely equivalent to the original model (\ref{eq:matrixising}) 
\cite{ez,johnston}.                            
This equivalence can be explicitly seen by
replacing
\begin{equation}
  \begin{array}{rcl}
  X & \rightarrow &  \displaystyle \frac{1}{ \sqrt{2}} ({U + V})
  \\
  Y & \rightarrow &  \displaystyle\frac{1}{ \sqrt{2}} ({U -V})
  \\
  \hat{g} &  \rightarrow &  \displaystyle {g}/{ \sqrt{2}}\ .
  \end{array}\labell{eq:xygmaps}
\end{equation}
Because of this correspondence, the partition functions on a closed
manifold for the two theories are precisely the same.  Nevertheless,
this does not by itself imply that Kramers-Wannier duality is
preserved in the presence of gravity; the transformation
(\ref{eq:xygmaps}) exchanges the disorder and spin operators, and
takes homogeneous $X$ boundary conditions in the dual model to free
boundary conditions in the original model (and vice versa).  A true
duality would not exchange operators and would relate amplitudes with
the same types of boundary conditions.  In \cite{cot2} we examine the
issue of duality between these models more closely.

\subsection{Generating equation}

We again study the generating function for a disk. Boundary conditions on
the disk are specified by a string of labels $x$ or $y$, whereas in the
Ising theory the labels are taken to live outside the boundary of the
disk.  Taking into account that there must be an even number of $y$'s
on the boundary (since each $y$ on the boundary is the endpoint of a 
loop), we can expand the generating function as
\begin{equation}
  \Phi(x,y) = 1+p_x x + p_{xx}x^2 + p_{yy}y^2 + p_{xxx}x^3
  +p_{xyy}(xyy + yxy + yyx) + \cdots \ ,
\end{equation}
where the coefficients may be thought of as expectation values:
\begin{equation}
  p_{w(x,y)} = {1\over N}\left\langle {\rm Tr}\; w(X,Y) \right\rangle \ .
  \labell{eq:relate2}
\end{equation}

We can now derive the generating equation for $\Phi$ using the geometric
techniques of the previous sections.  We obtain
\begin{equation}
  \Phi=1+{1\over 1-c}\Bigl[\hat{g} x\partial_x^2\Phi + \hat{g} x
  \partial_y^2\Phi + x\Phi x\Phi \Bigr]
  +{1\over 1+c}\Bigl[\hat{g} y\partial_y\partial_x\Phi +
  \hat{g} y\partial_x\partial_y\Phi + y\Phi y\Phi\Bigr]\ .
  \labell{eq:dualgen}
\end{equation}
The terms in (\ref{eq:dualgen}) correspond to moves that are analogous to
those for the Ising theory. Note, however, that the equation is not
symmetric in $x$ and $y$. Again there is a direct relation between
(\ref{eq:dualgen}) and the Schwinger-Dyson equations 
\begin{eqnarray}
 \partial_x\Phi&=&\displaystyle{1\over 1-c}\Bigl[\hat{g} \partial_x^2\Phi + 
  \hat{g} \partial_y^2\Phi + \Phi x\Phi \Bigr]\nonumber
\\
\partial_y\Phi&=&\displaystyle
  {1\over 1+c}\Bigl[\hat{g} \partial_y\partial_x\Phi +
  \hat{g} \partial_x\partial_y\Phi + \Phi y\Phi\Bigr]
\end{eqnarray}
for the matrix model.

As we did in the Ising theory, we may define functions which depend only 
on the single variable $x$:
\begin{equation}
  \phi_{w(x,y)}(x) = \left[w(\partial_x,\partial_y) \Phi\right]_{y=0}\ ,
\end{equation}
where $w(x,y)$ begins and ends with $y$. It is then possible to expand
$\Phi(x,y)$ as
\begin{equation}
\begin{array}{rcl}
  \Phi(x,y) &=& \phi(x)+ yy\phi_{yy}(x) + yxy\phi_{yxy}(x)
  \\
  &&\qquad +
  xyy\partial_x\phi_{yy}(x)+ yxxy\phi_{yxxy}(x)+yyyy\phi_{yyyy}(x)+ \cdots
\end{array}  
\end{equation}

Bearing in mind the asymmetry between $X$ and $Y$, we can use
(\ref{eq:dualgen}) to compute the generating function $\phi(x)$ for
homogeneous $X$ boundary conditions. As mentioned above, $\phi$ has two
interpretations: it is identically equal to the generating function for
free boundary conditions for the Ising theory,
while it can also be interpreted as
a generating function for homogeneous boundary conditions for the
dual Ising theory of (\ref{eq:matrixdual}).

It is straightfoward to
derive the closed set of five equations 
\begin{eqnarray}
  \phi &=& 1 + \displaystyle{1\over 1-c}\Bigl[\phi ^2\,x^2 + 
  \hat{g}x\partial_x^2\phi + 
   \hat{g}x\phi_{yy} \Bigr]\nonumber
  \\
  \phi_{yy} &=&  \displaystyle{1\over 1+c}\Bigl[\phi   + 
  \hat{g} \phi_{yxy} + \hat{g}\partial_x\phi_{yy}\Bigr ]\nonumber
  \\
  \phi_{yxy} &=& \displaystyle{1\over 1+c}\Bigl[p_x\phi+ \hat{g}\phi_{yxxy} +
  \hat{g}\partial_x\phi_{yxy}\Bigr]\labell{eq:d1d5}
  \\
  \phi_{yxy} &=& \displaystyle{1\over 1-c}\Bigl[\hat{g} \phi_{yxxy} + 
  \hat{g}\phi_{yyyy}\Bigr]\nonumber
  \\
  \partial_x\phi_{yy} &=& \displaystyle{1\over 1-c}\Bigl[x\phi\phi_{yy} + 
  \hat{g}\phi_{yyyy} +
  \hat{g}\partial_x^2\phi_{yy} \Bigr]\nonumber
\end{eqnarray}
from (\ref{eq:dualgen}). 
These equations can be combined to solve for $\phi$.
Once again replacing the derivatives of $\Phi$ with algebraic
expressions, we obtain a cubic equation for $\phi$:
\begin{equation}
  \hat{f}_3\phi^3 + \hat{f}_2\phi^2+\hat{f}_1\phi+\hat{f}_0 = 0
  \labell{eq:dualcubic}
\end{equation}
with coefficients
\begin{equation}
\begin{array}{rcl}
  \hat{f}_0 &=& - 4c{\hat{g}^2}(1+xp_x +x^2p_{xx}) + 2c\hat{g}x(3-c)(1+p_x x) 
  - 2c{x^2}(1-c^2) - {\hat{g}^2}{x^2}
  \\
  \hat{f}_1 &=&  4c{\hat{g}^2} - 2c\hat{g}x(3-c) + 2c{x^2}(1-c^2) +
  \hat{g}{x^3}(1 - 3c) - 
      2{\hat{g}^2}p_x{x^3}  
  \\
  \hat{f}_2 &=& x^2\left({\hat{g}^2} - \hat{g}{x}(1-5c) -
      2c{x^2}(1+c)\right)
  \\
  \hat{f}_3&=& \hat{g}x^5
  \end{array}
\end{equation}
As in the case of the Ising theory, it is straightforward, given an
expression for the homogenenous disk amplitude, to derive expressions for
amplitudes with nontrivial domain structure on the boundary through an
iterative procedure. In this case, however, the objects of interest are the
generating functions with a finite (even) number of $Y$'s on the boundary.
For example, the two-domain generating function $\tau(x_1,x_2)$ is defined
to be the generating function for all boundary configurations with a $Y$
(the starting point of the domain wall) followed by a series of $X$'s
followed by another $Y$ (the end point of the domain wall) followed by
another series of $X$'s.  It follows from (\ref{eq:dualgen}) that
$\tau(x_1, x_2)$ is given by the equation
\begin{equation}
  \tau(x_1,x_2)={1\over 1+c}\Bigl[\phi(x_1)\phi(x_2) +
  \hat{g}x_1^{-1}(\tau(x_1,x_2)-\phi_{yy}(x_2)) + \hat{g}x_2^{-1}
  (\tau(x_1,x_2)-\phi_{yy}(x_1))\Bigr]\ .
\end{equation}
An analysis of the continuum limit of the homogeneous and higher
domain amplitudes for the dual model is given in Ref. \cite{cot2}.

\section{General cubic two-matrix model}
\label{sec:general}

The previous two sections have described the Ising and dual Ising theories,
which correspond in the continuum limit to two distinct descriptions of the
$c = 1/2$ CFT coupled to quantum gravity.  In this section we investigate
the most general two-matrix model with cubic interactions.  This general
class of matrix models cannot be solved by saddle point or orthogonal 
polynomial methods, since the form of the interaction terms prevents the
angular degrees of freedom from being integrated out using the methods
of \cite{mehta,iz}.  However, by using the techniques of
this paper it is possible to find the homogeneous disk amplitude for any
model in this class.

After a description of the general model and a discussion of the
quartic equation satisfied by the homogeneous disk
amplitude in the general case, we look at a simple one-parameter family
of models which interpolate between the Ising and dual Ising theories.
Every model in this family has a homogeneous disk
amplitude which satisfies a quartic equation.
These homogeneous disk amplitudes can be interpreted as the disk
amplitude for the regular Ising theory coupled to gravity in the
presence of a boundary magnetic field.  This set of models is of
considerable intrinsic interest, and we shall study it in more detail
in a subsequent paper \cite{cot3}.

\subsection{Definition and generating equation}

The general two-matrix model with cubic interactions is specified
by its action, which we may write in the form
\begin{equation}
  S={1\over{2(\alpha \beta-\gamma^2)}}\Tr\left(\alpha \X^2
  +\beta \Y^2-2\gamma \X\Y\right)
  -{1\over 3}\Tr (a\X^3+3b\X^2\Y+3d\X\Y^2+e\Y^3)\ .
\labell{eq:ess}
\end{equation}
The coefficients of the quadratic terms have been chosen to yield a
simple form for the propagator, which in the $(\X, \Y)$ basis is
\begin{equation}
  \left(\matrix{\beta & \gamma \cr \gamma & \alpha \cr}\right)\ .
\end{equation}
Note that $\alpha$, which multiplies $\X^2$ in the action, is what
propagates $\Y$ to $\Y$, and vice-versa.

{}From the point of view of triangulations, this theory may be thought
of as one in which the {\it edges} of each triangle are labeled either
$\X$ or $\Y$; weights are assigned to each triangle according 
to the labels on its three edges and to each pair of neighboring
edges of triangles or boundaries.   Thus, the weight of a given
triangulation is
\begin{equation}
  W(\Delta) = {1\over{S(\Delta)}}N^{\chi(\Delta)}a^{qqq(\Delta)}
  b^{qqr(\Delta)}d^{qrr(\Delta)}e^{rrr(\Delta)}\alpha^{rr(\Delta)}
  \beta^{qq(\Delta)}\gamma^{qr(\Delta)}\ ,
\end{equation}
where $qqq(\Delta)$ is the number of triangles with three $\X$ edges,
$qq(\Delta)$ is the number of times an $\X$ edge is identified with
another $\X$ edge, and so on.  The partition function is given by
\begin{equation}
  Z = \sum_{\Delta} W (\Delta) = \int {\rm D} \X \; {\rm D} \Y \;
  \exp \left[-N S(\X,\Y)\right]\ .
\end{equation}

As in the previous sections, we can use geometric arguments of the kind
leading to (\ref{eq:generateising}) and (\ref{eq:dualgen}) in order to derive
an equation satisified by the generating function associated with a disk.
A boundary of length $l$ is specified by a string of $l$ labels $q$ or $r$;
however, these labels do not correspond to ``spins'' in any simple way.
Once again, a marked edge on the boundary is either attached to a triangle
or to another boundary edge.  Taking into account all possible combinations
of labelings for the marked edge and the neighboring triangle or boundary
edge, we find that the generating function $\Phi(q,r)$ satisfies
\begin{equation}\begin{array}{rcl}
  \Phi(q,r) &=& 1 + \beta q\Phi q\Phi + \gamma(q\Phi r\Phi + r\Phi q\Phi) 
  +\alpha r\Phi r\Phi
\\
  &&+(a'q +b'r)\partial_q^2\Phi +(b''q +d''r)
  (\partial_q\partial_r\Phi+\partial_r\partial_q\Phi)+
  (d'q +e'r)\partial_r^2\Phi\ , 
\end{array}\labell{eq:generalgenerate}\end{equation}
where we have defined the new
variables
\begin{equation}\begin{array}{rcl}
  a'&=&\beta a+\gamma b\ , \\
  b'&=&\alpha b+\gamma a \qquad\qquad b''=\beta b+\gamma d\ , \\
  d'&=&\beta d+\gamma e \qquad\qquad d''=\alpha d+\gamma b\ , \\
  e'&=&\alpha e+\gamma d\ .
\end{array}\labell{eq:reds}\end{equation}
The form of these new variables has a geometric interpretation.
As an example consider $d'$, the coefficient of $q\partial_r^2\Phi$.
This term originates from a marked edge labeled $q$ being attached
to a triangle, which when removed leaves two edges each labeled by
$r$.  In such a situation the triangle to be removed could have
been labeled either $qrr$ or $rrr$.  In the first case the weight is
given by multiplying the weight $d$ associated with an $qrr$ 
triangle by the weight $\beta$ which propagates $q$ to $q$, while in
the second case we multiply $e$ (for a $rrr$ triangle) by $\gamma $
(for an $q$-$r$ propagator).  It is therefore natural to define
$d'=\beta d+\gamma e$, and similarly for the rest of (\ref{eq:reds}).

It is straightforward to check that special cases of the generating
equation (\ref{eq:generalgenerate}) yield the previous equations
(\ref{eq:generateising}) for the Ising theory and (\ref{eq:dualgen}) for
the dual theory.

\subsection{Homogeneous boundary conditions}

As with the models studied in the previous sections, there are no general
techniques for solving the generating equation (\ref{eq:generalgenerate}),
but we can derive a polynomial equation for the generating function $\phi$
restricted to homogeneous boundary conditions (all $\X$'s on the boundary).
Unfortunately, in the general 2-matrix model the resulting equation for
$\phi$ is a quartic with extremely complicated coefficients, which it is
impractical to present explicitly.  Here we will describe the steps
followed in deriving the quartic.

We begin by expanding $\Phi$ as
\begin{equation}
\begin{array}{rcl}
  \Phi(q,r) &=& 1 +p_q q + p_r r + p_{qq}q^2 + p_{rr}r^2
  +p_{qr}(qr+rq) + p_{qqq}q^3 + p_{rrr}r^3
  \\  
  && \qquad +p_{qqr}(qqr+qrq+rqq) + p_{qrr}(qrr+rqr+rrq) + \ldots
\end{array}
\end{equation}
We then define generating functions which depend on $q$ alone,
as in the previous sections:
\begin{equation}
  \phi_{w(q,r)}(q) = \left[w(\partial_q,\partial_r) \Phi\right]_{r=0}\ ,
\end{equation}
To solve for $\phi$, we would like to find a set of
equations which closes on a set of these generating functions.  It
turns out that there are 10 equations which close on the 
quantities $(\phi,\phi_r,\phi_{rr},\phi_{rqr},\phi_{rrr},\phi_{rqqr},
\phi_{rqrr},\phi_{rrrr})$:
\begin{equation}
\begin{array}{rcl}
 \partial_q\phi &=&  \beta q \phi^2 +d'\phi_{rr} +a'\partial_q^2\phi 
   + 2b''\partial_q\phi_r
    \\
 \phi_r &=& \gamma  q \phi^2+ e'\phi_{rr} + b' \partial_q^2\phi 
 + 2 d'' \partial_q\phi_r
    \\
 \partial_q\phi_r &=& \gamma  \phi + \beta q\phi \phi_r + 
   b''\phi_{rqr} + d'\phi_{rrr} + a'\partial_q^2\phi_r 
   + b''\partial_q\phi_{rr}
    \\
 \partial_q^2\phi_r &=& \gamma  p_q \phi + \beta \phi_r 
   + \beta q\phi \partial_q\phi_r+ 
   b'' \phi_{rqqr} + d' \phi_{rqrr} + a' \partial_q^3\phi_r + 
      b'' \partial_q\phi_{rqr}
    \\
 \partial_q\phi_{rr}  &=& \gamma  p_r \phi + \gamma  \phi_r 
   + \beta q \phi \phi_{rr}+ 
   b'' \phi_{rqrr} + d' \phi_{rrrr} + a' \partial_q^2\phi_{rr} + 
      b'' \partial_q\phi_{rrr}
    \\
 \phi_{rr} &=& \alpha\phi +\gamma q \phi\phi_r + d'' \phi_{rqr}
   + e'\phi_{rrr} + b'\partial^2_q\phi_r +d''\partial_q\phi_{rr}
    \\
 \phi_{rqr} &=&  p_q \alpha \phi + \gamma  \phi_r+ d'' \phi_{rqqr} + 
   e' \phi_{rqrr} + \gamma  q \phi \partial_q\phi_r + b' \partial_q^3\phi_r + 
   d'' \partial_q\phi_{rqr}
    \\
 \phi_{rrr} &=& p_r \alpha \phi + \alpha \phi_r + \gamma  q \phi \phi_{rr} + 
   d'' \phi_{rqrr} + e' \phi_{rrrr} + b' \partial_q^2\phi_{rr} + 
   d'' \partial_q\phi_{rrr}
    \\
 \phi_{rqr} &=& 2\gamma\phi_r + \beta q\phi_r^2 + a'\phi_{rqqr} +
   2b''\phi_{rqrr} + d'\phi_{rrrr}
    \\
 \phi_{rrr}  &=& 2 \alpha \phi_r + \gamma  q \phi_r^2 + b' \phi_{rqqr} + 
   2 d'' \phi_{rqrr} + e' \phi_{rrrr} 
\end{array}
\labell{eq:gens}
\end{equation}
The first eight of these equations come from considering terms in
(\ref{eq:generalgenerate}) which begin with a certain string; for example
the third equation (for $\partial_q\phi_r$) comes from terms of the form
$qrqqq\ldots$.  The last two equations come from terms of the form
$qrqqq\ldots qqqr$ and $rrqqq\ldots qqqr$, which are related by cyclic 
symmetry to the functions $\phi_{rqr}$ and $\phi_{rrr}$, respectively.

These equations may be transformed into algebraic relations by replacing
the derivatives as in the previous two sections; the resulting system of
equations can be used to derive a single quartic equation involving only
$\phi$. 

Note that we can cut down somewhat on the number of undetermined
variables by looking at equations (\ref{eq:gens}) order by order.  These
give the following relations between the constants:
\begin{equation}\begin{array}{rcl}
  p_q &=& d'p_{rr} + a'p_{qq} + 2b'' p_{qr}\\
  p_{qq} &=& \beta + d'p_{qrr} + a'p_{qqq} + 2b''p_{qqr}\\
  p_r &=& e'p_{rr} + b'p_{qq} + 2d''p_{qr}\\
  p_{qr} &=& \gamma  + e'p_{qrr} + b'p_{qqq} + 2d''p_{qqr}\\
  p_{qr} &=& \gamma  + 2b''p_{qrr} + d'p_{rrr} + a'p_{qqr}\\
  p_{rr} &=& \alpha + 2d''p_{qrr} + e'p_{rrr} + b'p_{qqr}
\end{array}
\end{equation}
These allow $(p_{qq}, p_{rr}, p_{qqq}, p_{qqr},
p_{qrr}, p_{rrr})$ to be expressed in terms of the
three constants $(p_q, p_r, p_{qr})$, although we shall not present
the explicit results here.

\subsection{Models interpolating between Ising and dual models}

In Sections \ref{sec:ising} and \ref{sec:dual}, we discussed models in
which Ising spins were coupled directly to triangulations and their
duals.  These two geometrically distinct formulations of
the $c = 1/2$ theory are related by a simple coordinate
transformation, as was discussed in section \ref{sec:dual}.  In this
section, we explore a continuous 1-parameter family of models which
interpolate between the Ising and dual Ising theories.  This entire
family of theories are related to the original Ising theory by linear
transformations on the defining hermitian matrices.  The homogeneous
disk amplitude in these interpolating theories can be interpreted
physically as the disk amplitude in the original Ising theory where an
external magnetic field is imposed on the boundary.  Recently, there
has been much work on understanding the effects of boundary magnetic
fields on the Ising CFT in flat space \cite{gz,klm,chatt}; the
discussion here provides a formalism by which similar questions can be
asked of the theory after coupling to 2D gravity.  A more detailed
investigation of the physical properties of this theory will be
carried out in a separate publication \cite{cot3}.

The original Ising theory is defined by the action
\begin{equation}
  S={1\over{2}}\Tr (U^2+ V^2)-c \Tr UV
  -{g\over 3}\Tr (U^3+V^3)\ ,
\end{equation}
while the dual theory is defined by 
\begin{equation}
  S={{1-c}\over{2}}\Tr X^2 + {{1+c}\over{2}}\Tr Y^2
  -{\hat{g}\over 3}\Tr (X^3+3XY^2)\ .
\end{equation}
The two models are related by the transformation (\ref{eq:xygmaps}).
We now consider the more general continuous family of
theories defined by the following transformation of the variables
of the original Ising theory:
\begin{equation}\begin{array}{rcl}
  \X &=& \s U + \t V \ , \\
  \Y &=& \u U + \v V \ .
\end{array}
\end{equation}
We obtain a theory of the two matrices $\X$ and $\Y$, with action given by
(\ref{eq:ess}), where the parameters are expressed in terms of
$\s,\t,\u,\v$ and the couplings $c$ and $g$ of the original model:
\begin{eqnarray}
  \alpha &=& {1\over{1-c^2}}(\u^2 + \v^2 + 2c\u\v) \nonumber \\
  \beta &=& {1\over{1-c^2}}(\s^2 + \t^2 + 2c\s\t) \nonumber \\
  \gamma  &=& {1\over{1-c^2}}(\s\u + \t\v + c\s\t + c\u\v) \nonumber \\
  a &=& {{g}\over{(\s\v - \t\u)^3}}(\v^3 - \u^3)\labell{eq:vals} \\
  b &=& {{g}\over{(\s\v - \t\u)^3}}(\s\u^2 - \t\v^2) \nonumber \\
  d &=& {{g}\over{(\s\v - \t\u)^3}}(\t^2\v - \s^2\u) \nonumber \\
  e &=& {{g}\over{(\s\v - \t\u)^3}}(\s^3 - \t^3)  \nonumber
\end{eqnarray}
The original Ising theory corresponds to $\s=\v=1$, $\t=\u=0$, while
the dual Ising theory corresponds to $\s=\t=-\u=\v=1/\sqrt{2}$, with
the identification $\hat{g}=g/\sqrt{2}$.

The generating equation for this set of models may be derived from the
geometric arguments used in previous sections, or directly from
substitution of (\ref{eq:vals}) into (\ref{eq:generalgenerate}):
\begin{equation}
\begin{array}{rcl}
  \Phi(q,r) &=& 1 + \displaystyle {1\over{1-c^2}}\Bigl[
  (\s^2 + \t^2 +2c\s\t) q\Phi q\Phi 
  \\
  &&\qquad\qquad +(\s\u + \t\v +c\s\v + c\t\u) (q\Phi r\Phi + r\Phi q\Phi)  
  \\
  &&\qquad\qquad + (\u^2 + \v^2 +2c\u\v)r\Phi r\Phi\Bigr] 
  \\
  &&+\displaystyle {g\over{(1-c^2)(\s\v-\t\u)^2}}\Bigl[
  (c\s\u^2 + \t\u^2 + \s\v^2 + c\t\v^2)q\partial_q^2\Phi  
  \\
  &&\qquad\qquad+(\u + \v)(c\u^2 + \u\v - c\u\v + 
  c\v^2)r\partial_q^2\Phi 
  \\
  &&\qquad\qquad-(c\s^2\u + \s\t\u + \s\t\v + c\t^2\v)q
  (\partial_q\partial_r\Phi+\partial_r\partial_q\Phi) 
  \\
  &&\qquad\qquad-(c\s\u^2 + \s\u\v + \t\u\v + c\t\v^2)r
  (\partial_q\partial_r\Phi+\partial_r\partial_q\Phi) 
  \\
  &&\qquad\qquad+(\s + \t)(c\s^2 + \s\t - c\s\t + c\t^2)q 
  \partial_r^2\Phi 
  \\
  &&\qquad\qquad+(c\s^2\u + \t^2\u + \s^2\v + c\t^2\v)r\partial_r^2\Phi
  \Bigr] \ .
\end{array}
\end{equation}
In a similar fashion, we may substitute (\ref{eq:vals}) into the quartic
satisfied by the generating function $\phi$ for homogeneous boundary
conditions in the general 2-matrix model, to obtain a new quartic which
is somewhat simpler.  Since $\phi$ describes boundary
conditions which are all $\X$'s, we expect that the quartic will be
independent of $\u$ and $\v$ (which merely specify $\Y$ in terms of the
original variables $U$ and $V$), and this turns out to be the case.
We do not present the quartic here, but will examine it as part of
a discussion of the interpolating models in Ref.~\cite{cot3}.

The generating function $\phi(q)$
describes the amplitude for a disk with all $\X$'s on the boundary.  Since
$\X = \s U + \t V$, the coefficient of $q^n$ in the expansion for $\phi$
is the correlation function
\begin{equation}
  \left\langle {1\over N}\Tr (\s U + \t V)^n \right\rangle\ .
\end{equation}
We can therefore calculate the disk amplitude for the original
Ising theory with a family of different boundary conditions,
ranging from homogeneous ($\s=1$, $\t=0$) to completely free
($\s=\t$).  Writing
\begin{equation}\begin{array}{rcl}
\s & = & \alpha e^{h}\\
\t & = & \alpha  e^{-h}
\end{array}
\end{equation}
where $\alpha$ is a normalization constant,
we see that each choice of $\s,\t$ corresponds to a choice of boundary
magnetic field.
Homogeneous boundary conditions correspond to an infinite boundary
field, while free boundary conditions correspond to a vanishing
boundary field.  We shall discuss the continuum limit and interpretations
of these boundary conditions in Ref.~\cite{cot3}.

\section{3-state Potts model}

As mentioned previously, the approach to discretized gravity theories
via generating equations in free variables, when applied to matrix
model theories, may be thought of as an extension of the technique of
Schwinger-Dyson equations.  Although such equations are in principle
applicable to models where conventional methods (such as orthogonal
polynomials) are inadequate, in practice it has proven difficult to
take advantage of this greater generality.  One set of models for
which orthogonal polynomials seem to fail is those in which the target
space graph (defining the matter fields) contains a nontrivial cycle,
since the nontrivial cycle prevents simultaneous diagonalization of
the matrices \cite{mehta,iz}.  There is some evidence that loop
equation methods also fail for these models,\footnote{We would like to
thank Matthias Staudacher for explaining this issue to us.} although
to our knowledge a clear understanding of why this failure occurs is
still lacking.

The simplest theory with a nontrivial cycle is the 3-state Potts
model.  In fact this model is solvable by conventional matrix model
techniques, as there exists a change of variables which allows the
matrices to be diagonalized \cite{kazakov3,kostov3,daul}; presumably
the same change of variables would allow solution by the methods
considered here.  Nevertheless, in order to illustrate the
difficulties encountered when a nontrivial cycle is present, in this
section we examine the 3-state Potts model in its original form, and
discuss the obstacles to finding a solution for the homogeneous disk
amplitude.

In the 3-state Potts model we associate to each triangle one of
three ``spins,'' labelled by the free variables $x_0$, $x_1$ and
$x_2$.  There is a nearest-neighbor interaction which is the same
for any two neighboring unequal spins.  (The target graph may
therefore be thought of as a triangle with the three free variables
labelling the vertices.)
The weight associated with each triangulation $\Delta$ is
\begin{equation}
  W(\Delta) = \frac{1}{S (\Delta)}  N^{\chi (\Delta)} 
  g^{n (\Delta)} c^{\nu(\Delta)}  \ ,
  \labell{eq:3spweight}
\end{equation}
where $\nu(\Delta)$ is the total number of edges which connect 
unlike spins.  The partition function may be cast in matrix model
form, in which case it becomes an integral over three matrices
$X_0,X_1,X_2$, with an action given by
\begin{equation}\begin{array}{rcl}
  S &=& \frac{1}{2(1+c-2c^2)}\Tr\left[(1+c)(X_0^2+X_1^2+X_2^2)
  -2c(X_0X_1+X_0X_2+X_1X_2)\right] \\
  && \qquad +\frac{g}{3}\Tr(X_0^3+X_1^3+X_2^3)\ .
  \end{array}\labell{eq:3spact}
\end{equation}
Once again we have chosen the quadratic terms in the action such
that the propagator takes on a simple form.

{}From the geometric arguments familiar from previous sections,
we derive a generating equation governing the disk amplitude
in this model:
\begin{equation}\begin{array}{rcl}
  \Phi(x_0,x_1,x_2) &=& 1 + x_0\Phi(x_0+cx_1+cx_2)\Phi 
  + x_1\Phi(cx_0+x_1+cx_2)\Phi \\ 
  && \quad +x_2\Phi(cx_0+cx_1+x_2)\Phi +g(x_0+cx_1+cx_2)\partial^2_0\Phi \\
  && \quad +g(cx_0+x_1+cx_2)\partial^2_1\Phi 
  +g(cx_0+cx_1+x_2)\partial^2_2\Phi \ .
  \end{array}\labell{eq:3spgen}
\end{equation}
Continuing to follow the procedure outlined in previous sections, 
{}from (\ref{eq:3spgen}) we derive a set of seven equations
in seven functions $(\phi, \phi_1, \phi_{11}, \phi_{111}, \phi_{122},
\phi_{1111}, \phi_{1122})$ of the single variable $x_0$:
\begin{equation}\begin{array}{rcl}
  \phi &=& 1 + 2gcx_0\phi_{11} + gx_0\partial_0^2\phi + x_0^2\phi^2 \\
  \phi_1 &=& g(1+c)\phi_{11} + gc\partial_0^2\phi + cx_0\phi^2 \\
  \phi_{11} &=& gc\phi_{122} + g\phi_{111} + gc\partial^2_0\phi_1 
    + cx_0\phi\phi_1 + \phi \\
  \partial_0\phi_1 &=& gc\phi_{122} + gc\phi_{111} +g\partial^2_0\phi_1
    +x_0\phi\phi_1 + c\phi \\
  \phi_{111} &=& gc\phi_{1122} + g\phi_{1111}  + gc\partial^2_0\phi_{11}
    +\phi_1 + p_1\phi + cx_0\phi\phi_{11} \\
  \phi_{122} &=& g\phi_{1122} + gc\phi_{1111} + gc\partial^2_0\phi_{11}
    + c\phi_1 + cp_1\phi + cx_0\phi\phi_{11} \\
  \partial_0\phi_{11} &=& gc\phi_{1122} + gc\phi_{1111} 
    + g\partial_0^2\phi_{11} + c\phi_1 + cp_1\phi + x_0\phi\phi_{11}
  \end{array}\labell{eq:3sp7}
\end{equation}
We might expect to be able to solve these equations to get a single
equation for $\phi(x_0)$.  However, closer inspection reveals that
they are not independent, with one redundancy hidden among them.

It is possible to uncover a geometric understanding of why these equations
are redundant.  Two of the equations -- the one for $\phi_1$ and the one
for $\partial_0\phi_1$ -- can be thought of as governing what happens when
an edge is removed from a disk with a single $x_1$ and a number of $x_0$'s;
the first comes from removing the $x_1$ edge, and the second from removing
the $x_0$ edge immediately next to the $x_1$.  We could imagine acting on
such a disk by removing both the $x_1$ edge and the $x_0$ next to it, in
either order; the resulting expressions must be compatible, and must be
equivalent to performing the appropriate operations on the right hand sides
of the two equations.  But the result of performing such operations on the
right hand sides is already contained in the other equations in this set
(\ref{eq:3sp7}), so there must be a consistency condition, reducing the
number of independent equations by one. On the other hand, the two
equations for $\phi_{11}$ and $\partial_0\phi_{11}$ do not share this
problem, since the right hand side of the equation for
$\partial_0\phi_{11}$ contains $\phi_{1111}$ and $\phi_{1122}$, and the act
of removing an edge from the triangulations represented by these quantities
is not contained in the seven equations.  We therefore are left with six
equations in seven unknowns, insufficient to find a single equation for
$\phi$.

By expanding the set of unknowns to include the three additional
quantities ($\phi_{11111}$, $\phi_{11211}$, $\phi_{11222}$), we can
derive four additional equations:
\begin{equation}\begin{array}{rcl}
  \phi_{1111} &=& gc\phi_{11222} + g\phi_{11111} 
    + gc\partial^2_0\phi_{111} + \phi_{11} + p_1\phi_1 + p_{11}\phi
    + cx_0\phi\phi_{111} \\
  \phi_{1122} &=& gc\phi_{11211} + g\phi_{11222}
    + gc\partial_0^2\phi_{122} + \phi_{11} + cp_1\phi_1 
    + cp_{12}\phi + cx_0\phi\phi_{122} \\
  \partial_0\phi_{111} &=& gc\phi_{11222} + gc\phi_{11111}
    + g\partial^2_0\phi_{111} + c\phi_{11} + cp_1\phi_1
    + cp_{11}\phi + x_0\phi\phi_{111} \\
  \partial_0\phi_{122} &=& gc\phi_{11211} + gc\phi_{11222}
    + g\partial^2_0\phi_{122} + c\phi_{11} + cp_1\phi_1
    + cp_{12}\phi + x_0\phi\phi_{122}
  \end{array}\labell{eq:3sp4}
\end{equation}
But once again, there is an additional consistency condition which
reduces the number of independent equations by one.  In this case,
we have introduced equations with $\phi_{1111}$ and $\phi_{1122}$
on the left hand sides, which leads to another condition on the
moves represented by the $\phi_{11}$ and $\partial_0\phi_{11}$
equations, as discussed above.  

We are therefore still left with one fewer equation than unknowns.  As far
as we can determine, this will continue to be the case no matter how many
additional equations we introduce, although we have not been able to find a
general proof of this fact.  It remains an open question whether some other
method of solution can yield information about this model from the
generating equation (\ref{eq:3spgen}).

\section{Multiple Ising spins}

The major limitation on the usefulness of matrix models is the
fact that no known methods for solving matrix models apply to theories
with central charge $c > 1$.  Since the systems of the greatest
physical interest fall into this category, the ``$c = 1$ wall'' has
always been a major obstacle in the further understanding of 2D
gravity and string theories.  However, the fact that it is difficult
to integrate out angular degrees of freedom in matrix models with $c >
1$ has always seemed to be a rather technical problem, and it has
been unclear whether some other method might
make it possible to find analytic results or solutions for systems
with larger central charge.

In this section we shall discuss a class of matrix models formed by taking
$k$ independent Ising spins coupled to quantum gravity.  The model with $k$
spins corresponds to a CFT with $c = k/2$ coupled to gravity, and can be
described as a $2^k$-matrix model.  We analyze the 4-matrix model
corresponding to the 2-spin $c = 1$  
CFT coupled to gravity.  As was the case with the 3-state Potts model, we
find that it seems to be impossible to find a closed system of equations
describing the disk amplitude.

\subsection{Multiple spin model}

The multiple spin Ising theory is defined in a manner precisely
analogous to the Ising theory in Section \ref{sec:ising}, except that
each triangle carries $k$ independent Ising spins.  Thus, in the
language of triangulations, we sum over all triangulations $\Delta$
where each triangle carries $k$ spins $\sigma_1, \ldots, \sigma_k$.
We set the interaction energy to be identical for each type of spin,
so that the weight of a given triangulation $\Delta$ is given by
\begin{equation}
  W (\Delta) =
  \frac{1}{S (\Delta)}  N^{\chi (\Delta)} g^{n (\Delta)}
  c^{\nu(\Delta)}
\end{equation}
where $\nu(\Delta)$ is the total number of adjacent
pairs of unequal spins with the same index, and $c$ is the interaction
between unequal spins (note that we the normalization factor $1-c^2$
is moved into the matrix model action for convenience in this section).

The matrix model description of this partition function is given by an
integral over $2^k$ matrices $X_{\sigma_1 \cdots \sigma_k}$, where
$\sigma_i \in\{+1,-1\}$:
\begin{equation}
  Z = \sum_{\Delta} W (\Delta)
  = \int \prod_{\{\sigma_1, \ldots, \sigma_k\}} {\rm D}  X_{\sigma_1
  \cdots \sigma_k} \; 
  \exp \left(-N S \right)
\end{equation}
with
\begin{equation}
  S = \sum_{\{\sigma_1, \ldots, \sigma_k\}} \left[ \frac{1}{2(1-c^2)^k}
  \sum_{\{\tau_1, \ldots, \tau_k\}} 
(-c)^{{1\over 2}\left(k-\sum_{i=1}^k \sigma_i\tau_i\right)}
 \Tr(X_{\sigma_1 \cdots \sigma_k} X_{\tau_1 \cdots \tau_k})
  -\frac{g}{3} \Tr(X_{\sigma_1 \cdots \sigma_k}^3)\right]\ .
  \labell{eq:msmact}
\end{equation}
By the same geometric argument as has been used throughout the paper, we
can write a generating equation for the disk amplitude as a function of
free noncommuting variables $x_{\sigma_1\cdots\sigma_k}$ which takes the
very simple form
\begin{equation}
  \Phi = 1 +\sum_{\sigma_1, \ldots,\sigma_{k}, \tau_1, \ldots, \tau_k}
  c^{{1\over 2}\left(k-\sum_{i=1}^k \sigma_i\tau_i\right)}
  x_{\sigma_1\cdots\sigma_k} \left(\Phi
  x_{\tau_{1}\cdots \tau_{k}}\Phi +
  g\partial^2_{\tau_{1}\cdots \tau_{k}}\Phi\right)\ .
  \labell{eq:generaldisk}
\end{equation}

\subsection{2-spin model}

For simplicity, we shall restrict attention for the remainder of this
section to the 2-spin model with $c = 1$.  This 4-matrix model has a
continuum limit which can be identified with the orbifold $c = 1$
bosonic string at radius $r = 1$ \cite{ginsparg,bfgm}.    Labeling the 4
matrices according to their binary representations for shorthand
({\it i.e.}, $X_3 = X_{+ +}$, $X_2 = X_{+ -}$, $X_1 = X_{-+}$,  
$X_0 = X_{--}$), the matrix model action is given by
\begin{equation}
  S = \frac{1}{2} \sum_{i,j = 0}^{3} \left[ C_{ij} \Tr
  X_iX_j\right] - \frac{g}{3} \sum_{i=0}^3 \Tr X_i^3 \ ,
\end{equation}
with
\begin{equation}
C_{ij} = \frac{1}{(1-c^2)^2}\left(
  \matrix{1 & -c & -c & c^2 \cr
  -c & 1 & c^2 & -c \cr
  -c & c^2 & 1 & -c \cr
  c^2 & -c & -c & 1 \cr}\right)\ .
  \labell{eq:cinv}
\end{equation}
The generating equation (\ref{eq:generaldisk}) becomes
\begin{equation}\begin{array}{rcl}
  \Phi & = & 1+ x_{3}\Phi x_{3}\Phi+ x_{2}\Phi x_{2}\Phi
  + x_{1}\Phi x_{1}\Phi+ x_{0}\Phi x_{0}\Phi  \\
  && +c\left((x_{2}+x_{1})\Phi(x_{3}+x_{0})\Phi 
  +(x_{3}+x_{0})\Phi (x_{2}+x_{1})\Phi\right)  \\
  && +c^2\left(x_{3}\Phi x_{0}\Phi+ x_{2}\Phi x_{1}\Phi 
  + x_{1}\Phi x_{2}\Phi+ x_{0}\Phi x_{3}\Phi\right) \\
  &&+g\left(x_{3}\partial^2_{3}\Phi +x_{2}\partial^2_{2}\Phi
  +x_{1}\partial^2_{1}\Phi +x_{0}\partial^2_{0}\Phi\right) \\
  &&+gc\left((x_{3}+x_{0})(\partial^2_{2}\Phi+\partial^2_{1}\Phi)
  +(x_{2}+x_{1})(\partial^2_{3}\Phi+\partial^2_{0}\Phi)\right)\\
  &&+gc^2\left(x_{3}\partial^2_{0}\Phi +x_{2}\partial^2_{1}\Phi
  +x_{1}\partial^2_{2}\Phi +x_{0}\partial^2_{3}\Phi\right).
\end{array}
\labell{eq:generating2}
\end{equation}
As for the 3-state Potts model, it appears not to be possible to
derive a single algebraic equation for the homogeneous disk amplitude
from this equation. Once again, however, we do not have any proof that
this must be the case.  However, it seems unlikely that any algebraic
equation for the homogeneous disk amplitude exists in view of the
logarithmic behavior of other previously studied $c = 1$ models. 

It is nevertheless conceivable that some other means of analyzing the
generating equation (\ref{eq:generating2}) could lead to a single equation
characterizing $\phi$.  Furthermore, we mention the promising result that
all coefficients $p_{s}$ can be determined in terms of the three
coefficients $p_{0}$, $p_{000}$, $p_{0000}$, just as in the Ising  theory
where only two such coefficients were necessary.  Presumably, however, any
method for solving the $c = 1$ model would not generalize to the $c = k/2$
models with $k > 2$ since the number of unknowns in the higher $k$ models
grows much faster than when $k = 2$.

\section{Dually weighted models}

So far in this paper, all of the specific models which we have
considered have simple descriptions in terms of the matrix model
formalism (although we described a class of boundary conditions for
the Ising theory in
Section~\ref{sec:boundarycorrelation} which cannot easily be described
using matrix model methods) .  However, the methodology we have
developed is applicable to a much wider class of discrete gravity
models.  In general, we can construct discrete matter fields by
endowing a dynamical triangulation with an arbitrary set of local
discrete data.  For example, we can simultaneously give discrete
labels ($f_i$, $e_i$, $v_i$) to faces (polygons), edges, and vertices
of a triangulation.  If we make the weight function local, in the
sense that the only interactions are between adjacent faces, edges,
and vertices, then the partition function corresponding to such a
theory is of the form
\begin{eqnarray}
Z & = &  \sum_{\Delta, \Lambda}\frac{1}{S (\Delta)}  N^{\chi (\Delta)}
\left[ (\prod_{f \in {\rm faces}} \exp (\rho (f,\{e \in \partial f\},
\{v \in \partial f\})))\right.  \nonumber\\
 & & \hspace{1in}\cdot
(\prod_{e \in {\rm edges}} \exp (\eta (\{f:e \in \partial f\},e,
\{v \in \partial e\})))\label{eq:general}\\
& & \hspace{1in}\cdot\left.
(\prod_{v \in {\rm vertices}} \exp (\nu (\{f:v \in \partial f\},
\{e:v \in \partial e\},v))) \right]\nonumber
\end{eqnarray}
where $\Lambda$ is  a set of combinatorial data $(f_i,e_i,v_i)$ on the
simplices in a triangulation $\Delta$;  $\rho$ is a function of the
geometry and labels of a fixed face $f$ and the edges and vertices on
its boundary; similarly $\eta$ is a function of a fixed edge $e$, the
faces of which $e$ is a boundary, and the vertices on the boundary of
$v$; and $\nu$ is a function of a vertex $v$ and the edges and faces
of which it is a boundary.  

A generating equation can be derived for a model of the general form
(\ref{eq:general}) by associating independent noncommuting variables
with the discrete labels on edges, vertices, and faces.  Although the
complexity of such an equation will increase significantly as the set
of discrete matter fields increases, in principle the methods of this
paper can provide  an approach to the analysis of any such theory.

A particular set of theories in this more general class are the
``dually weighted graph'' models recently studied by Di~Francesco
and Itzykson \cite{dfi} and by Kazakov,
Staudacher, and Wynter \cite{ksw}.  In these models, weights are
associated with vertices of a triangulation according to the
coordination number, and weights are associated with the
faces of a triangulation according to the number of incident edges as
in a general one-matrix model.  The dually weighted graph models are
of particular interest because they contain continuous families of
theories which interpolate between conformal field theories on flat
space and conformal field theories completely coupled to Liouville
gravity.  Although it is possible to construct most dually weighted
graph models as matrix models by introducing external matrices
encoding the vertex weights, the resulting models are rather unwieldy
and difficult to solve by standard matrix model techniques.  An example of
a model which can be analyzed using matrix model methods is the model
where vertices of both the original and dual graph are restricted to
have even coordination number \cite{ksw}.

We now derive the generating equation for a general dually weighted
graph model.  The most general dually weighted graph model has a
weight of $g_i$ for every face with $i$ edges, and a weight of $w_i$
for every vertex with coordination number $i$.  Thus, the partition
function for a general dually weighted graph model is given by
\begin{equation}
Z = \sum_{\Delta} \frac{1}{S (\Delta)}  N^{\chi (\Delta)}
\left[ \prod_{i}
g_i^{F_i (\Delta)}
w_i^{V_i (\Delta)} \right]
\end{equation}
where $F_i (\Delta)$ is the number of $i$-gon faces in $\Delta$ and
$V_i (\Delta)$ is the number of vertices with
coordination number $i$.

\begin{figure}
\centerline{
\psfig{figure=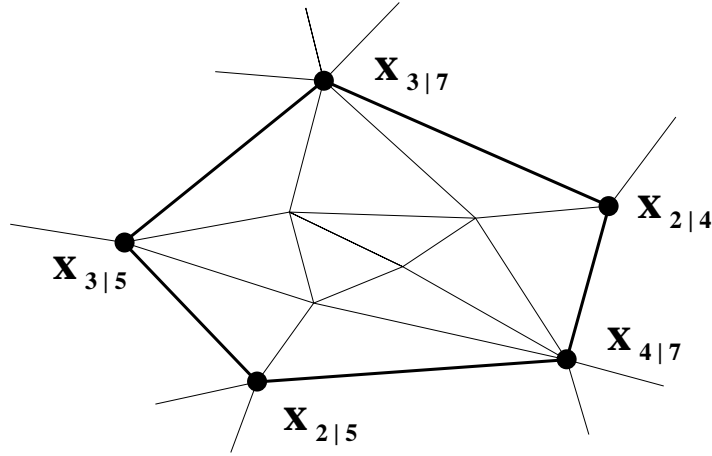,angle=0,height=6cm}}
\begin{center}
\parbox{5in}{
\caption{\em Variables $x_{i | j}$ on the vertices around the boundary
of a disk in a dually weighted graph model.  The subscript $j$ indicates
the coordination number of the vertex; $i$ indicates the number of
faces incident on the vertex on the interior of the disk.}  \label{f:dwg}
}
\end{center}
\end{figure}

To describe the generating function for the disk amplitude of this model,
we must associate noncommuting variables with the vertices on the boundary
of the disk.  These noncommuting variables must contain information about
the coordination number of the vertex, as well as the number of faces
inside the disk which are incident on that vertex.  Thus, we choose
variables $x_{i | j}$, with $i < j$ to denote a vertex with $i$ interior
faces and a coordination number of $j$.  (An example is shown in
Figure~\ref{f:dwg}.)

The generating equation for the general dually weighted graph model is
\begin{eqnarray}
\Phi & = &  1 + \left(\sum_{i < j,i' < j',k}
g_k x_{i + 1 | j} x_{ i' + 1 | j'} \partial_{i' | j'}
\partial_{*}^{k-2} \partial_{i |j} \Phi \right) \\
& &\hspace{0.6in} +
\left(\sum_{i + j \geq k,i' + j' \geq k'}
x_{i | k} x_{i' | k'} (\partial_{i' + j' -k'| k'} \Phi)
x_{j'  | k'} x_{j | k}(\partial_{i + j -k| k} \Phi) \right)
\end{eqnarray}
where we have defined
\begin{equation}
\partial_* = \sum_{i > 2} w_i \partial_{i -1 | i}.
\end{equation}

As an extremely simple example, let us consider a ``flat space''
dually weighted graph model where only hexagonal faces and vertices with
coordination number 3 are allowed ({\em i.e.}, $g_i = \delta_{i,6}$; $w_i =
\delta_{i, 3}$).  This is simply a theory of ``triangulations'' of
flat surfaces by hexagons; each term in the disk amplitude for this
theory corresponds to a closed path on the honeycomb lattice which
bounds an embedded oriented surface composed of hexagons.  
The generating equation for this model reduces to 
(writing $x_i = x_{i | 3}$)
\begin{eqnarray*}
\Phi & = &  1 +x_2 x_2 \partial_1 \partial_2^4 \partial_1 \Phi
+x_2 x_1 \partial_0 \partial_2^4 \partial_1 \Phi 
+x_1 x_2 \partial_1 \partial_2^4 \partial_0 \Phi
+x_1 x_1 \partial_0 \partial_2^4 \partial_0 \Phi \\
& &+
x_2 \left[ x_2 (\partial_1 \Phi) x_2 +
x_1 (\partial_0 \Phi) x_2 + x_2 (\partial_0 \Phi) x_1 \right] x_2
(\partial_1 \Phi) \\
& &+
x_2 \left[ x_2 (\partial_1 \Phi) x_2 +
x_1 (\partial_0 \Phi) x_2 + x_2 (\partial_0 \Phi) x_1 \right] x_1
(\partial_0 \Phi) \\
& &+
x_1 \left[ x_2 (\partial_1 \Phi) x_2 +
x_1 (\partial_0 \Phi) x_2 + x_2 (\partial_0 \Phi) x_1 \right] x_2
(\partial_0 \Phi).
\end{eqnarray*}

There are a number of slightly more interesting models which are very
simple to describe, and which connect flat space theories such as the above
with fluctuating 2D gravity theories which presumably correspond to
coupling in some form of matter fields.  For example, consider the theory
with only square faces, and vertices of coordination number between 3 and 5
defined by
\begin{eqnarray}
g_i & = &  \delta_{i,4}\nonumber\\
w_i & = &  \delta_{i,4} + a \delta_{i,3} + b \delta_{i,5}. \label{eq:squares}
\end{eqnarray}
When  $a = b = 0$, this is simply another flat space theory.  As $a$
and $b$ become nonzero, fluctuations are introduced.  It should be
possible by tuning the ratio $a/b$ to find a 1-parameter family of
theories which correspond to a continuous limit; any detailed
understanding of the behavior of this class of theories would be
extremely interesting for understanding the effects of coupling
dynamical gravity to a  flat theory.  The generating equation for this
model is  simply the restriction of (\ref{eq:general}) to the
theory with weights (\ref{eq:squares}).  

It would be interesting to see whether for these simple models there is any
reasonable algebraic method for extracting information from the generating
equation.

\section{Conclusion}

We have developed a systematic method for analyzing discretized
theories of 2D gravity with matter.  Our method combines a
geometric/combinatorial approach with the tools of free variables and
loop equations which have previously been used to study matrix models.
Although this approach gives rise to equations which are essentially
equivalent to the Schwinger-Dyson loop equations for standard matrix
model theories, our results are more general in several ways.  First,
because our methodology is completely based on a geometric definition
of the discrete gravity theories, the analysis applies to models which
cannot be described in terms of the matrix model language.  Second, we
can describe correlation functions in standard matrix model theories
using variables which are apparently inaccessible using the matrix
model formalism.  For example, in \cite{cot2} we describe the
calculation using our techniques of the exact correlation function
between disorder operators in the Ising theory.  As far as we have
been able to ascertain, this correlation function cannot be computed
in any simple way using matrix model methods.  Finally, we believe
that the principal result presented in this paper, the general
single generating equation for the generating function of all disk
amplitudes, represents a conceptual and algebraic simplification and
synthesis of previous work in this direction.

The methods we have presented here have enabled us to analyze in a
systematic fashion a variety of  models describing 2D gravity with
matter fields.  One question which has never been fully resolved is
the question of which matrix models allow solution by loop equation
methods.  To date, the only theories which have proven tractable using
this approach have been the pure gravity and $c = 1/2$ models.  We
have found that it is possible to use this approach to derive
polynomial equations satisfied by the generating function of disk
amplitudes not only in the standard Ising formulation of the $c = 1/2$
theory, but also in the dual formulation, and even in a continuous
family of models connecting these two formulations.  In addition, 
we have found that a very general class of two-matrix
models, which {\em cannot} be solved by  orthogonal polynomials, have
uniform boundary condition disk amplitude generating functions which 
satisfy a quartic.

We have attempted to solve several other models using these methods,
including the 3-state Potts model and the $c = 1$ model defined by two
Ising spins coupled to gravity.  We have found that these models
resist solution, in accordance with the empirical belief that models
whose target graphs contain loops somehow have more intrinsic degrees
of freedom, and are not soluble by loop equations or algebraic means.
However, in our attempt to solve these equations, we uncovered
algebraic structure to the obstruction, characterized by degeneracies
in subsets of the loop equations.   This algebraic structure may give
some hint of how to  use other approaches to solve these models.

In the discrete theories that we have considered, the existence of a 
single equation describing the effect of all possible moves that reduce a
triangulation suggests that this equation should be related to a Hamiltonian
operator for the theory. In this context it would be interesting to 
consider the continuum limit of the equation in terms of noncommuting 
variables and see if can be related to recent attempts to construct 
string field hamiltonians for theories with $c<1$ [22-30]. 

The approach we have detailed here simplifies the
calculation of correlation functions in a variety of interesting 2D
theories.  Some particular results along these lines will be described
in \cite{cot2,cot3}.  However,   the major conceptual issue which
needs to be dealt with for  these methods to be applicable in a truly
general context is the development of some more sophisticated and
powerful methods for dealing with algebraic equations in noncommuting
variables such as (\ref{eq:generateising}).  This is a very difficult
problem, as has been emphasized in a number of recent papers
\cite{voiculescu,singer,gg,dl}; however, in these works and
others some progress has been made,  giving hope that in the future
these equations may seem more tractable.

\section*{Acknowledgments}
We would like to thank M. Douglas, D. Gross, I. Harrus, A. Matytsin,
G. Moore, I.M. Singer, and M. Staudacher for helpful conversations.  
This work was
supported in part by the National Science Foundation under grants
PHY/9200687 and PHY/9108311, by the U.S. Department of Energy (D.O.E.)
under cooperative agreement DE-FC02-94ER40818, by the divisions of 
applied mathematics of the D.O.E. under contracts DE-FG02-88ER25065
and DE-FG02-88ER25065, and by the European Community
Human Capital Mobility programme.

\end{document}